\documentclass{article}
\usepackage{slashed,verbatim,subfigure}
\usepackage[numbers,sort&compress]{natbib}
\usepackage{amsmath}
\usepackage{amssymb}
\usepackage{appendix}
\usepackage{graphicx}
\usepackage{dcolumn}
\usepackage{authblk}
\usepackage{bm}
\usepackage[colorlinks=true,linktocpage=true,
linkcolor=blue,citecolor=blue]{hyperref}
\usepackage[a4paper]{geometry}
\newcommand{\be}{\begin{eqnarray}}
\newcommand{\ee}{\end{eqnarray}}
\begin{document}
\large
\title{\bf{Deciphering the viscous properties and the Bjorken expansion of the QGP medium at finite angular velocity}}
\author[1,2]{Shubhalaxmi Rath\thanks{shubhalaxmirath@gmail.com}}
\author[1]{Nicol\'{a}s A. Neill\thanks{naneill@outlook.com}}
\affil[1]{Centro Multidisciplinario de F\'isica, Vicerrector\'ia de Investigaci\'on, Universidad Mayor, 8580745 Santiago, Chile}
\affil[2]{Instituto de Alta Investigaci\'{o}n, Universidad de Tarapac\'{a}, Casilla 7D, Arica, Chile}
\date{}
\maketitle
\begin{abstract}
We have studied the viscous properties as well as the Bjorken expansion of a rotating QGP medium. In the noncentral events of heavy-ion collisions, the produced medium can carry a finite angular momentum 
with a finite range of angular velocity. This rotation can significantly affect various properties, including viscous properties and the expansion of the QGP medium. Using a novel relaxation time approximation for the collision integral in the relativistic Boltzmann transport equation at finite angular velocity, we have calculated the shear and bulk viscosities and compared them with their counterparts 
in the standard relaxation time approximation within the kinetic theory approach. Our results show that the angular velocity increases both shear and bulk viscosities, suggesting an enhanced momentum transfer within the medium and greater fluctuations in local pressure. This rotational effect on viscosities is more evident at lower temperatures than at higher temperatures. Our analysis also shows that, compared to the standard relaxation time approximation, the shear viscosity is lower while the bulk viscosity is higher in the novel relaxation time approximation for all temperatures. Additionally, some observables related to the flow characteristic, fluid behavior and conformal symmetry of the medium are markedly impacted due to rotation. We have also studied the hydrodynamic evolution of matter within the Bjorken boost-invariant scenario and have found that the energy density evolves faster in the presence of finite rotation than in the nonrotating case. Consequently, rapid rotation accelerates the cooling process of the QGP medium. 

\end{abstract}

\newpage

\section{Introduction}
The deconfined state of matter, {\em i.e.}, quark-gluon plasma (QGP) can be affected by various extreme conditions in ultrarelativistic heavy-ion collisions at the Relativistic Heavy Ion Collider (RHIC) 
and the Large Hadron Collider (LHC). There is evidence that the noncentral heavy-ion collisions could generate a large angular momentum varying between $10^3 \hbar$ and $10^5 \hbar$, with an angular velocity in the range of 0.01 GeV to 0.1 GeV or even more, and a significant portion of this angular momentum is retained in the produced fireball \cite{Deng:PRC93'2016,Jiang:PRC94'2016,Wang:PRD99'2019,Wei:PRD105'2022}.
A larger impact parameter ($b$) and higher center-of-mass energy ($\sqrt{s_{NN}}$) may result in a greater angular momentum of the QGP. The conservation of total angular momentum contributes to the longevity of this large rotational motion. Thus, similar to the strong magnetic fields, high temperatures, and large chemical potential, rapid rotation can be considered as one of the extreme conditions that may significantly alter various properties of the QGP medium. This extreme condition is also found in spinning neutron stars \cite{Berti;MNRAS358'2005} and nonrelativistic bosonic cold atoms \cite{Fetter:RMP81'2009}, in addition to QGP medium. The rapid rotation of the QGP can have significant effects in heavy-ion collisions, including the global polarization of thermal photons, dileptons, and final hadrons with spin, as well as the enhancement of the elliptic flow and broadening of the transverse momentum spectra \cite{Liang:PLB629'2005,Becattini:PRC77'2008}. There also exist some collective effects of the strong magnetic field and the rapid rotation in heavy-ion collisions, such as the chiral vortical effect \cite{Son:PRL103'2009,Kharzeev:PRL106'2011}, the chiral vortical wave \cite{Jiang:PRD92'2015}, the chiral magnetic effect \cite{Fukushima:PRD78'2008,Kharzeev:NPA803'2008}, the chiral magnetic wave \cite{Kharzeev:PRD83'2011,Burnier:PRL107'2011}, the spin alignment of vector mesons \cite{Liang:PLB629'2005}, the emission of circularly polarized photons \cite{Ipp:PLB666'2008} and the spin polarization of observed particles in heavy-ion collisions \cite{Betz:PRC76'2007,Becattini:AP323'2008}. How different extreme conditions modulate the properties of the QGP medium remains an active area of research in the high energy physics community. 

In general, angular momentum manifests as vorticity in the system. The polarization of emitted hadrons in heavy-ion collisions provides a signature of vorticity in the produced matter. The degree of such 
polarization depends mainly on the initial energy and the impact parameter. For a review of the current status of the vortical behavior of QGP, see ref. \cite{Becattini:ARNPS70'2020}. Rotational effects in heavy-ion collisions can significantly modify the thermodynamics and the critical temperature of the QGP-to-hadronic phase transition. Several models predict a decrease in the critical temperature due to rotation, including the Nambu-Jona-Lasinio model \cite{Chernodub:JHEP1701'2017,Wang:PRD99'2019}, the local density approximation with boundary condition \cite{Ebihara:PLB764'2017}, the hadron resonance gas model \cite{Fujimoto:PLB816'2021} 
and the holographic QCD approach \cite{Chen:JHEP2107'2021}. On the other hand, lattice simulations \cite{Braguta:JETPL112'2020,Braguta:PRD103'2021} have reported an increase in the critical temperature due to the rotation of the QGP medium. Studies using the Nambu-Jona-Lasinio model have shown that rapid rotation lowers the critical temperature for chiral symmetry breaking by suppressing the chiral condensate in the rotating medium \cite{Jiang:PRL117'2016}. In recent years, noticeable progress has been made in the study of magnetized and rotating QCD medium. For instance, the properties of rotating free fermion states were studied in refs. \cite{Iyer:PRD26'1982,Becattini:AP323'2008,Becattini:PRD84'2011,Victor:PLB734'2014,Victor:PRD93'2016}, the system of interacting fermions were explored using effective field theory methods in refs. \cite{Chen:PRD93'2016,Jiang:PRL117'2016,Ebihara:PLB764'2017}, using the holographic approach in refs. \cite{McInnes:NPB887'2014,McInnes:NPB911'2016} and using the Nambu-Jona-Lasinio model in ref. \cite{Chernodub:JHEP1701'2017}. Additionally, ref. \cite{Singh:PRD100'2019} reported that the presence of vorticity could suppress the dilepton yield. Different studies have also examined the properties of the magnetized thermal QCD medium using various methods \cite{Rath:JHEP1712'2017,Bandyopadhyay:PRD100'2019,Rath:EPJA55'2019,Karmakar:PRD99'2019,Hattori:PRD94'2016,
Feng:PRD96'2017,Fukushima:PRL120'2018,Rath:PRD100'2019,Rath:EPJC80'2020,Kurian:EPJC79'2019,
Rath:EPJA59'2023,Nam:PRD87'2013,Hattori:PRD96'2017,Li:PRD97'2018,Denicol:PRD98'2018,Rath:PRD102'2020,
Rath:EPJC81'2021,Rath:EPJC82'2022,Hees:PRC84'2011,Shen:PRC89'2014,Tuchin:PRC88'2013,Mamo:JHEP1308'2013,
Fukushima:PRD93'2016}. In a recent study \cite{Rath:EPJC85'2025}, the transport coefficients related to charge and heat for a rotating QGP medium have been explored using the novel relaxation time approximation \cite{Rocha:PRL127'2021} within the kinetic theory approach. 

In this work, we study the impact of rotation on the viscous properties and expansion dynamics of the matter produced in ultrarelativistic heavy-ion collisions. The shear viscosity ($\eta$) and the bulk viscosity ($\zeta$) play an important role in the hydrodynamic description of QGP. In general, shear viscosity describes the momentum transfer within the medium, while bulk viscosity is related to the change in local pressure due to fluid expansion or compression. Thus the resistance to any deformation at constant volume is measured by the shear viscosity, whereas the bulk viscosity is associated with the resistance to change in volume at constant shape. These viscosities can modulate different observables in heavy-ion collisions, including the elliptic flow coefficient and the hadron transverse momentum spectrum. The ratio of shear viscosity to entropy density ($\eta/s$) can provide essential information about the fluid behavior of matter, while the ratio of bulk viscosity to entropy density ($\zeta/s$) is an indicator of conformal symmetry in the system.
Moreover, these ratios play a crucial role in identifying the phase transition of QGP \cite{Csernai:PRL97'2006}. The determination of shear and bulk viscosities is essential for interpreting experimental data on particle multiplicities, transverse momentum spectra and elliptic flow \cite{Ryu:PRL115'2015}. Studies based on Anti-de Sitter/Conformal Field Theory (AdS/CFT) have established a lower bound on $\eta/s$ at $1/(4\pi)$, characterizing QGP as a strongly interacting perfect fluid \cite{Kovtun:PRL94'2005}, while perturbative QCD calculations predict a comparatively higher value of $\eta/s$ \cite{Arnold:PRD74'2006}. Further, the value of $\zeta/s$ is very small and approaches zero in systems that exhibit conformal symmetry. Lattice QCD simulations \cite{Borsonyi:JHEP11'2010,Bazavov:PRD90'2014} have reported a peak in the trace anomaly near the critical temperature, indicating that QCD does not show conformal symmetry in this regime. The presence of a strong magnetic field significantly modifies both shear and bulk viscosities. The components of shear and bulk viscosities along the direction of magnetic field exist only in the strong magnetic field regime \cite{Lifshitz:BOOK'1981,Tuchin:JPG39'2012,Hattori:PRD96'2017}.
Similarly, rapid rotation can significantly alter the values of shear and bulk viscosities.
Therefore, this work systematically explores the effects of rotation on these transport coefficients.  
This study aims to quantify the deviation of a rotating medium from ideal hydrodynamic behavior. Additionally, we explore the impact of rapid rotation on observables pertaining to the flow characteristics, fluid behavior, conformal symmetry and expansion dynamics of the QGP medium. The shear and bulk viscous coefficients are computed by solving the relativistic Boltzmann transport equation 
using the novel relaxation time approximation \cite{Rocha:PRL127'2021} within the kinetic theory approach. 
In the standard relaxation time approximation (RTA), there may be violations of the local particle number conservation and energy-momentum conservation laws while using the momentum-dependent relaxation time. However, this difficulty can be cured in the novel relaxation time approximation, where the collision integral takes a modified form to ensure the preservation of fundamental microscopic conservation laws \cite{Cercignani:BOOK'2002,Rocha:PRL127'2021}. Furthermore, in this work, the rotational effects are encoded in the angular velocity-dependent parton distribution functions. Parton interactions are 
modeled through their thermal masses within the quasiparticle model, where QGP is considered to be a medium composed of thermally massive quasiparticles. 

The matter produced in the ultrarelativistic heavy-ion collisions immediately expands hydrodynamically and this expansion can be significantly affected by various extreme conditions.
The impact of the magnetic field on the expansion of QGP has already been studied by various research groups.
For instance, the $(1+1)$-dimensional expansion has been analyzed within the Bjorken picture at finite magnetic field by comparing the energy densities associated with the magnetic field and the fluid on an event-by-event basis \cite{Umut:PRC89'2014,Roy:PRC92'2015}. A study based on $(3+1)$-dimensional anomalous relativistic hydrodynamics \cite{arXiv:1412.0311} in the presence of a magnetic field has shown that the chiral magnetic effect exists in the charge-dependent hadron azimuthal correlations. The influence of a spatially inhomogeneous magnetic field on the anisotropic expansion in $(3+1)$-dimensional ideal hydrodynamics has been explored in ref. \cite{Pang:PRC93'2016}. Hydrodynamic expansion in the presence of a magnetic field has been studied by parameterizing the equation of state with finite magnetization in refs. \cite{Pu:PRD93'2016,Roy:PRC96'2017}. The effect of a strong magnetic field on the expansion of hot QCD matter has been examined using a paramagnetic equation of state in ref. \cite{Rath:EPJA55'2019}. This study indicates that the energy density evolves faster in the presence of a strong magnetic field, leading to a faster cooling of the medium. Additionally, the influence of shear and bulk viscosities on hydrodynamic expansion has been explored in the $(1+1)$-dimensional boost invariant picture in refs. \cite{Kajantie:NPA418'1984,Baym:NPA418'1984,Hosoya:NPB250'1985,Muronga:PRL88'2002}. To the best of our knowledge, no prior studies about the effect of rotation on the expansion properties of QGP are available 
in the literature. Therefore, in this work, we explore the effect of rotation on the expansion of QGP through an angular velocity-dependent hydrodynamic equation of motion. We observe that the energy density evolves faster in a rotating medium as compared to a nonrotating medium. As a result, rapid rotation speeds up the cooling of matter. 

This work is organized as follows. Section 2 is devoted to the calculations of the shear and bulk viscosities for a rotating QGP medium. The corresponding results are discussed in section 3. Section 4 explores various phenomena and observables associated with the viscous coefficients, such as the flow characteristic, the specific shear viscosity, the specific bulk viscosity and the Bjorken expansion of the rotating matter. Finally, section 5 presents our conclusions. 

\section{Shear and bulk viscous properties in rotating QGP medium}
Various properties of the QGP medium, including shear viscosity and bulk viscosity, are likely to be modulated by extreme conditions produced in ultrarelativistic heavy-ion collisions. One such condition is the rapid rotation of the medium. The rotation of the QGP medium can significantly affect the phase-space distribution functions of quarks and gluons. The partition function in such rotating system is given by
\be\label{P.F.}
Z=Tr\left[e^{\beta\left(-\hat{H}+\mu\hat{N}+\omega\hat{J}\right)}P_V\right]
,\ee
where $\omega$, $\hat{H}$, $\hat{N}$ and $\hat{J}$ represent the angular velocity, the Hamiltonian 
operator, the number operator and the angular momentum operator, respectively. 
Here, $T=\beta^{-1}$ and $P_V$ is the projector onto the localized states $|h_V\rangle$, {\em i.e.} $P_V=\sum_{h_V}|h_V\rangle\langle h_V|$, which forms a complete set of quantum states for the system in a finite region $V$. For a high-temperature QGP medium, considering thermally massive partons with spin $S$ in the Boltzmann limit, the partition function \eqref{P.F.} can be written as a product of single particle partition functions. Thus, the calculation of the partition function \eqref{P.F.} depends on the matrix elements of operators $\hat{h}$, $\hat{n}$ and $\hat{j}$ for single particles \cite{Becattini:AP325'2010} as
\be\label{M.E.}
\langle p,\tau|e^{\beta\left(-\hat{h}+\mu\hat{n}+\omega\hat{j}\right)}P_V|p,\sigma \rangle=\left(e^{\beta\left(u_\mu p^\mu+a\right)}+b\right)^{-1}\langle p,\tau|e^{\beta\omega\hat{j}}P_V|p,\sigma \rangle
,\ee
where $a=-\mu$, $+\mu$ and $0$ for quarks, antiquarks and gluons, respectively, $b=+1$ for quarks as well as for antiquarks, and $b=-1$ for gluons. Here, $\tau$ and $\sigma$ represent the polarization states. 
Following ref. \cite{Becattini:AP325'2010}, the distribution functions for quarks, antiquarks, and gluons in a rotating QGP medium can be expressed as
\be\label{Q.D.F.}
f_q &=& f^q_0\chi_q(\beta\omega), \\ 
\label{A.D.F.}\bar{f_q} &=& \bar{f^q_0}\chi_{\bar{q}}(\beta\omega), \\ 
\label{G.D.F.}f_g &=& f^g_0\chi_g(\beta\omega)
,\ee
where $\chi_{q,\bar{q},g}(\beta\omega)$ is the angular velocity dependent factor, given by 
\be\label{A.V.P.}
\chi_{q,\bar{q},g}(\beta\omega) &=& \frac{\sinh\left(\left(S_{q,\bar{q},g}+\frac{1}{2}\right)\beta\omega\right)}{\sinh\left(\frac{\beta\omega}{2}\right)}
,\ee
with $S_{q,\bar{q},g}$ representing the spin of the corresponding particle. The different quantities appearing in equations \eqref{Q.D.F.}, \eqref{A.D.F.} and \eqref{G.D.F.} are defined as follows, 
\begin{eqnarray}
&&\nonumber f^q_0=\frac{1}{e^{\beta\left(u_\mu p^\mu-\mu\right)}+1}, ~~ \bar{f^q_0}=\frac{1}{e^{\beta\left(u_\mu p^\mu+\mu\right)}+1}, ~~ f^g_0=\frac{1}{e^{\beta u_\mu p^\mu}-1}, \\  &&\nonumber \chi_q(\beta\omega)=\frac{\sinh(\beta\omega)}{\sinh(\beta\omega/2)}, ~~ \chi_{\bar{q}}(\beta\omega)=\frac{\sinh(\beta\omega)}{\sinh(\beta\omega/2)}, ~~ \chi_g(\beta\omega)=\frac{\sinh(3\beta\omega/2)}{\sinh(\beta\omega/2)} 
.\end{eqnarray}
Here, $\mu$ denotes the chemical potential for each quark flavor, $p_\mu\equiv\left(\omega_f,\mathbf{p}\right)$ for quarks and antiquarks, $p_\mu\equiv\left(\omega_g,\mathbf{p}\right)$ for gluons, where $\omega_f$ is the energy of the $f$th flavor quark or antiquark and $\omega_g$ is the energy of the gluon in the rotating QGP medium. In this work, the flavor degrees of freedom are taken into account explicitly. Further, the chemical potentials for flavors, $u$, $d$ and $s$ are considered to be the same ($\mu_f=\mu$). In the study of shear viscosity, bulk viscosity and related observables in a rotating QGP medium, we use the angular velocity dependent parton distribution functions given in equations \eqref{Q.D.F.}, \eqref{A.D.F.} and \eqref{G.D.F.}. 

When an external force disturbs the system, the energy-momentum tensor and the particle distribution functions get shifted infinitesimally from their equilibrium values. This shift can be expressed as  $T^{\mu\nu}\rightarrow {T^\prime}^{\mu\nu}=T^{\mu\nu}+\Delta T^{\mu\nu}$, $f_q\rightarrow f_q^\prime=f_q+\delta f_q$, $\bar{f}_q\rightarrow \bar{f}_q^\prime=\bar{f}_q+\delta \bar{f}_q$ and $f_g\rightarrow f_g^\prime=f_g+\delta f_g$, where $\Delta T^{\mu\nu}$ is the nonequilibrium part of the energy-momentum tensor, and $\delta f_q$, $\delta \bar{f}_q$ and $\delta f_g$ are the infinitesimal changes in quark, antiquark and gluon distribution functions, respectively. The energy-momentum tensor for a slightly nonequilibrium medium is expressed as
\be\label{N.E.M.}
{T^\prime}^{\mu\nu}=\int\frac{d^3{\rm p}}{(2\pi)^3}p^\mu p^\nu \left[\sum_f g_f\frac{\left(f_q^\prime+\bar{f}_q^\prime\right)}{{\omega_f}}+g_g\frac{f_g^\prime}{\omega_g}\right]
,\ee
where the subscript $f$ denotes quark flavor. In eq. \eqref{N.E.M.}, $g_f$ and $g_g$ represent the degeneracy factors for quarks and gluons, respectively. Similarly, we can write $\Delta T^{\mu\nu}$ as
\be\label{em1}
\Delta T^{\mu\nu}=\int\frac{d^3{\rm p}}{(2\pi)^3}p^\mu p^\nu \left[\sum_f g_f\frac{\left(\delta f_q+\delta \bar{f}_q\right)}{{\omega_f}}+g_g\frac{\delta f_g}{\omega_g}\right]
.\ee
We need to obtain the infinitesimal changes of the parton distribution functions appearing in eq. \eqref{em1} in order to calculate the shear and bulk viscosities. To calculate these infinitesimal changes, we solve the relativistic Boltzmann transport equation (RBTE) using a novel relaxation time approximation method within the kinetic theory approach, recently proposed in ref. \cite{Rocha:PRL127'2021}. For comparison, we also solve the RBTE using the relaxation time approximation method, and calculate the shear and bulk viscosities. 

\noindent{\bf\underline{Relaxation time approximation method:}} 

For the calculation of shear and bulk viscous coefficients, one can exclude the electromagnetic field part from the relativistic Boltzmann transport equation. In the relaxation time approximation method, the relativistic Boltzmann transport equations for quarks, antiquarks and gluons are written as
\be\label{R.B.T.E.Q (1)}
p^\mu\partial_\mu f_q^\prime=-\frac{p_\nu u^\nu}{t_f}\delta f_q, \\ 
\label{R.B.T.E.A (1)}p^\mu\partial_\mu \bar{f}_q^\prime=-\frac{p_\nu u^\nu}{t_{\bar{f}}}\delta \bar{f}_q, \\ 
\label{R.B.T.E.G (1)}p^\mu\partial_\mu f_g^\prime = -\frac{p_\nu u^\nu}{t_g}\delta f_g
,\ee
respectively, where the relaxation times for quark (antiquark), $t_f$ ($t_{\bar{f}}$) and gluon, $t_g$ are momentum-independent, whose forms are given \cite{Hosoya:NPB250'1985} by
\be\label{R.T.Q}
t_{f(\bar{f})} &=& \frac{1}{5.1T\alpha_s^2\log\left(1/\alpha_s\right)\left[1+0.12 
(2N_f+1)\right]} ~, \\ 
\label{R.T.G}t_g &=& \frac{1}{22.5T\alpha_s^2\log\left(1/\alpha_s\right)\left[1+0.06N_f\right]}
~.\ee
From equations \eqref{R.B.T.E.Q (1)}, \eqref{R.B.T.E.A (1)} and \eqref{R.B.T.E.G (1)}, the infinitesimal changes in the parton distribution functions are obtained as
\be
\delta f_q=-\frac{t_f p^\mu\partial_\mu f_q^\prime}{p_\nu u^\nu}, \\ 
\delta \bar{f}_q=-\frac{t_{\bar{f}} p^\mu\partial_\mu \bar{f}_q^\prime}{p_\nu u^\nu}, \\ 
\delta f_g=-\frac{t_g p^\mu\partial_\mu f_g^\prime}{p_\nu u^\nu}
.\ee
Utilizing the values of $\delta f_q$, $\delta \bar{f}_q$ and $\delta f_g$ in eq. \eqref{em1}, we get 
\be\label{em2 (1)}
\Delta T^{\mu\nu}=-\int\frac{d^3{\rm p}}{(2\pi)^3}\frac{p^\mu p^\nu}{p_\nu u^\nu} \left[\sum_f g_f\left(\frac{t_f p^\mu\partial_\mu f_q^\prime}{\omega_f}+\frac{t_{\bar{f}} p^\mu\partial_\mu \bar{f}_q^\prime}{\omega_f}\right)+g_g\frac{t_g p^\mu\partial_\mu f_g^\prime}{\omega_g}\right]
.\ee
Here, the partial derivative is given by $\partial_\mu=u_\mu D+\nabla_\mu$, with $D=u^\mu\partial_\mu$. In the local rest frame, the distribution functions can be expanded in terms of the gradients of temperature and flow velocity. The partial derivatives of the nonequilibrium quark, antiquark and gluon distribution 
functions are respectively determined as
\begin{eqnarray}
\nonumber\partial_\mu f_q^\prime = \beta f^q_0\left(1-f^q_0\right)\chi_q(\beta\omega)\left[u_\alpha p^\alpha u_\mu\frac{DT}{T}+u_\alpha p^\alpha\frac{\nabla_\mu T}{T}-u_\mu p^\alpha Du_\alpha\right. \\ \left.-p^\alpha\nabla_\mu u_\alpha+T\partial_\mu\left(\frac{\mu}{T}\right)\right]+\beta f^q_0\sinh\left(\frac{\beta\omega}{2}\right)T\partial_\mu\left(\frac{\omega}{T}\right)
,\end{eqnarray}
\begin{eqnarray}
\nonumber\partial_\mu \bar{f}_q^\prime = \beta \bar{f^q_0}\left(1-\bar{f^q_0}\right)\chi_{\bar{q}}(\beta\omega)\left[u_\alpha p^\alpha u_\mu\frac{DT}{T}+u_\alpha p^\alpha\frac{\nabla_\mu T}{T}-u_\mu p^\alpha Du_\alpha\right. \\ \left.-p^\alpha\nabla_\mu u_\alpha-T\partial_\mu\left(\frac{\mu}{T}\right)\right]+\beta \bar{f^q_0}\sinh\left(\frac{\beta\omega}{2}\right)T\partial_\mu\left(\frac{\omega}{T}\right)
,\end{eqnarray}
\begin{eqnarray}
\nonumber\partial_\mu f_g^\prime = \beta f^g_0\left(1+f^g_0\right)\chi_g(\beta\omega)\left[u_\alpha p^\alpha u_\mu\frac{DT}{T}+u_\alpha p^\alpha\frac{\nabla_\mu T}{T}-u_\mu p^\alpha Du_\alpha-p^\alpha\nabla_\mu u_\alpha\right] \\ +2\beta f^g_0\sinh\left(\beta\omega\right)T\partial_\mu\left(\frac{\omega}{T}\right)
.\end{eqnarray}
Using the expressions for $\partial_\mu f_q^\prime$, $\partial_\mu \bar{f}_q^\prime$ and $\partial_\mu f_g^\prime$ in eq. \eqref{em2 (1)} and substituting $\frac{DT}{T}=-\left(\frac{\partial P}{\partial \varepsilon}\right)\nabla_\alpha u^\alpha$ and $Du_\alpha=\frac{\nabla_\alpha P}{\varepsilon+P}$ from the energy-momentum conservation, we have
\be\label{em3 (1)}
\nonumber\Delta T^{\mu\nu} &=& \sum_f g_f\int\frac{d^3{\rm p}}{(2\pi)^3}\frac{p^\mu p^\nu \beta}{\omega_f}\left[\left\lbrace{t_f f^q_0\left(1-f^q_0\right)\chi_q(\beta\omega)}\left(\omega_f\left(\frac{\partial P}{\partial \varepsilon}\right)\nabla_\alpha u^\alpha\right.\right.\right. \\ && \left.\left.\left.\nonumber+p^\alpha\left(\frac{\nabla_\alpha P}{\varepsilon+P}-\frac{\nabla_\alpha T}{T}\right)-\frac{Tp^\alpha}{\omega_f}\partial_\alpha\left(\frac{\mu}{T}\right)+\frac{p^\alpha p^\beta}{\omega_f}\nabla_\alpha u_\beta\right)\right.\right. \\ && \left.\left.\nonumber -\frac{f^q_0 t_f}{\omega_f}\sinh\left(\frac{\beta\omega}{2}\right)Tp^\alpha\partial_\alpha\left(\frac{\omega}{T}\right)\right\rbrace\right. \\ && \left.\nonumber+\left\lbrace{t_{\bar{f}} \bar{f^q_0}\left(1-\bar{f^q_0}\right)\chi_{\bar{q}}(\beta\omega)}\left(\omega_f\left(\frac{\partial P}{\partial \varepsilon}\right)\nabla_\alpha u^\alpha\right.\right.\right. \\ && \left.\left.\left.\nonumber+p^\alpha\left(\frac{\nabla_\alpha P}{\varepsilon+P}-\frac{\nabla_\alpha T}{T}\right)+\frac{Tp^\alpha}{\omega_f}\partial_\alpha\left(\frac{\mu}{T}\right)+\frac{p^\alpha p^\beta}{\omega_f}\nabla_\alpha u_\beta\right)\right.\right. \\ && \left.\left.\nonumber -\frac{\bar{f^q_0} t_{\bar{f}}}{\omega_f}\sinh\left(\frac{\beta\omega}{2}\right)Tp^\alpha\partial_\alpha\left(\frac{\omega}{T}\right)\right\rbrace\right] \\ && \nonumber+g_g\int\frac{d^3{\rm p}}{(2\pi)^3}\frac{p^\mu p^\nu \beta}{\omega_g}\left[{t_g f^g_0\left(1+f^g_0\right)\chi_g(\beta\omega)}\left\lbrace\omega_g\left(\frac{\partial P}{\partial \varepsilon}\right)\nabla_\alpha u^\alpha\right.\right. \\ && \left.\left.+p^\alpha\left(\frac{\nabla_\alpha P}{\varepsilon+P}-\frac{\nabla_\alpha T}{T}\right)+\frac{p^\alpha p^\beta}{\omega_g}\nabla_\alpha u_\beta\right\rbrace-\frac{2f^g_0 t_g}{\omega_g}\sinh\left(\beta\omega\right)Tp^\alpha\partial_\alpha\left(\frac{\omega}{T}\right)\right]
.\ee
The pressure and the energy density can be determined from the energy-momentum tensor as $P=-\Delta_{\mu\nu}T^{\mu\nu}/3$ and $\varepsilon=u_\mu T^{\mu\nu}u_\nu$, respectively, where $\Delta_{\mu\nu}=g_{\mu\nu}-u_\mu u_\nu$ is the projection tensor. In the local rest frame, since $\Delta T^{00}=0$, only the spatial 
components of $\Delta T^{\mu\nu}$ depend on the velocity gradient. From eq. \eqref{em3 (1)}, the spatial components of $\Delta T^{\mu\nu}$ are written as
\be\label{em4 (1)}
\nonumber\Delta T^{ij} &=& \sum_f g_f\int\frac{d^3{\rm p}}{(2\pi)^3}\frac{p^i p^j \beta}{\omega_f}\left[\left\lbrace{t_f f^q_0\left(1-f^q_0\right)\chi_q(\beta\omega)}\left(\left(\omega_f\left(\frac{\partial P}{\partial \varepsilon}\right)-\frac{\rm p^2}{3\omega_f}\right)\partial_l u^l\right.\right.\right. \\ && \left.\left.\left.\nonumber -\frac{p^kp^l}{2\omega_f}W_{kl}-\frac{Tp^k}{\omega_f}\partial_k\left(\frac{\mu}{T}\right)+p^k\left(\frac{\partial_k P}{\varepsilon+P}-\frac{\partial_k T}{T} \right)\right)\right.\right. \\ && \left.\left.\nonumber -\frac{f^q_0 t_f}{\omega_f}\sinh\left(\frac{\beta\omega}{2}\right)Tp^k\partial_k\left(\frac{\omega}{T}\right)\right\rbrace\right. \\ && \left.\nonumber+\left\lbrace{t_{\bar{f}} \bar{f^q_0}\left(1-\bar{f^q_0}\right)\chi_{\bar{q}}(\beta\omega)}\left(\left(\omega_f\left(\frac{\partial P}{\partial \varepsilon}\right)-\frac{\rm p^2}{3\omega_f}\right)\partial_l u^l\right.\right.\right. \\ && \left.\left.\left.\nonumber -\frac{p^kp^l}{2\omega_f}W_{kl}+\frac{Tp^k}{\omega_f}\partial_k\left(\frac{\mu}{T}\right)+p^k\left(\frac{\partial_k P}{\varepsilon+P}-\frac{\partial_k T}{T}\right)\right)\right.\right. \\ && \left.\left.\nonumber -\frac{\bar{f^q_0} t_{\bar{f}}}{\omega_f}\sinh\left(\frac{\beta\omega}{2}\right)Tp^k\partial_k\left(\frac{\omega}{T}\right)\right\rbrace\right] \\ && \nonumber+g_g\int\frac{d^3{\rm p}}{(2\pi)^3}\frac{p^i p^j \beta}{\omega_g}\left[{t_g f^g_0\left(1+f^g_0\right)\chi_g(\beta\omega)}\left\lbrace\left(\omega_g\left(\frac{\partial P}{\partial \varepsilon}\right)-\frac{\rm p^2}{3\omega_g}\right)\partial_l u^l\right.\right. \\ && \left.\left.-\frac{p^kp^l}{2\omega_g}W_{kl}+p^k\left(\frac{\partial_k P}{\varepsilon+P}-\frac{\partial_k T}{T}\right)\right\rbrace-\frac{2f^g_0 t_g}{\omega_g}\sinh\left(\beta\omega\right)Tp^k\partial_k\left(\frac{\omega}{T}\right)\right]
.\ee
Here, we have used $\partial_k u_l=-\frac{1}{2}W_{kl}-\frac{1}{3}\delta_{kl}\partial_j u^j$ and $W_{kl}=\partial_k u_l+\partial_l u_k-\frac{2}{3}\delta_{kl}\partial_j u^j$. The shear and bulk viscosities 
are given by the coefficients of the traceless and trace parts of the dissipative contribution of the 
energy-momentum tensor, respectively. The spatial component of the nonequilibrium part of the 
energy-momentum tensor in the first order theory is 
defined \cite{Lifshitz:BOOK'1981,Hosoya:NPB250'1985,Landau:BOOK'1987} as
\be\label{definition}
\Delta T^{ij}=-\eta W^{ij}-\zeta\delta^{ij}\partial_l u^l
.\ee
Now, comparing equations \eqref{em4 (1)} and \eqref{definition}, we get the shear viscosity for a rotating QGP medium in the relaxation time approximation as
\begin{eqnarray}\label{iso.eta (1)}
\nonumber\eta &=& \frac{\beta}{30\pi^2}\sum_f g_f \int d{\rm p}~\frac{{\rm p}^6}{\omega^2_f}\left[t_f f^q_0\left(1-f^q_0\right)\chi_q(\beta\omega)+t_{\bar{f}}\bar{f^q_0}\left(1-\bar{f^q_0}\right)\chi_{\bar{q}}(\beta\omega)\right] \\ && +\frac{\beta}{30\pi^2} g_g \int d{\rm p}~\frac{{\rm p}^6}{\omega^2_g} ~ t_g f^g_0\left(1+f^g_0\right)\chi_g(\beta\omega)
~.\end{eqnarray}
Similarly, the bulk viscosity is obtained after comparing equations \eqref{em4 (1)} and \eqref{definition} as
\begin{eqnarray}\label{iso.zeta1 (1)}
\nonumber\zeta &=& \frac{1}{3}\sum_f g_f \int\frac{d^3{\rm p}}{(2\pi)^3}~\frac{{\rm p}^2}{\omega_f}\left[f^q_0\left(1-f^q_0\right)\chi_q(\beta\omega)A_f+\bar{f^q_0}\left(1-\bar{f^q_0}\right)\chi_{\bar{q}}(\beta\omega)\bar{A}_f\right] \\ && +\frac{1}{3}g_g \int\frac{d^3{\rm p}}{(2\pi)^3}~\frac{{\rm p}^2}{\omega_g}f^g_0\left(1+f^g_0\right)\chi_g(\beta\omega)A_g
~,\end{eqnarray}
where $A_f={\beta t_f}\left[\frac{{\rm p}^2}{3\omega_f}-\left(\frac{\partial P}{\partial \varepsilon}\right)\omega_f\right]$, $\bar{A}_f={\beta t_{\bar{f}}}\left[\frac{{\rm p}^2}{3\omega_f}-\left(\frac{\partial P}{\partial \varepsilon}\right)\omega_f\right]$ and $A_g={\beta t_g}\left[\frac{{\rm p}^2}{3\omega_g}-\left(\frac{\partial P}{\partial \varepsilon}\right)\omega_g\right]$. In the local rest frame, for the validity of the Landau-Lifshitz condition ($\Delta T^{00}=0$), the factors $A_f$, $\bar{A}_f$ and $A_g$ are replaced by $A_f\rightarrow A_f^\prime=A_f-b_f\omega_f$, $\bar{A}_f\rightarrow \bar{A}_f^\prime=\bar{A}_f-\bar{b}_f\omega_f$ and $A_g\rightarrow A_g^\prime=A_g-b_g\omega_g$, where $b_f$, $\bar{b}_f$ and $b_g$ are arbitrary constants related to the particle number and energy conservations. Using the ``00'' component of eq. \eqref{em3 (1)}, the Landau-Lifshitz conditions for $A_f$, $\bar{A}_f$ and $A_g$ are written as
\begin{eqnarray}
&&\sum_f g_f\int\frac{d^3{\rm p}}{(2\pi)^3} ~ \omega_f f^q_0\left(1-f^q_0\right)\chi_q(\beta\omega)\left(A_f-b_f\omega_f\right)=0 \label{A_i (1)} ~,~ \\ 
&&\sum_f g_f\int\frac{d^3{\rm p}}{(2\pi)^3} ~ \omega_f \bar{f^q_0}\left(1-\bar{f^q_0}\right)\chi_{\bar{q}}(\beta\omega)\left(\bar{A}_f-\bar{b}_f\omega_f\right)=0 \label{A_i.1 (1)} ~,~ \\ 
&&g_g\int\frac{d^3{\rm p}}{(2\pi)^3} ~ \omega_g f^g_0\left(1+f^g_0\right)\chi_g(\beta\omega)\left(A_g-b_g\omega_g\right)=0 \label{A_g (1)}
~,\end{eqnarray}
respectively. The quantities $b_f$, $\bar{b}_f$ and $b_g$ are determined by solving equations \eqref{A_i (1)}, \eqref{A_i.1 (1)} and \eqref{A_g (1)}. After solving, the expressions of these quantities are obtained 
as
\begin{eqnarray}
&&b_f=\frac{\sum_f g_f\int\frac{d^3{\rm p}}{(2\pi)^3} ~ \omega_f f^q_0\left(1-f^q_0\right)\chi_q(\beta\omega)A_f}{\sum_f g_f\int\frac{d^3{\rm p}}{(2\pi)^3} ~ \omega^2_f f^q_0\left(1-f^q_0\right)\chi_q(\beta\omega)}, \\ 
&&\bar{b}_f=\frac{\sum_f g_f\int\frac{d^3{\rm p}}{(2\pi)^3} ~ \omega_f \bar{f^q_0}\left(1-\bar{f^q_0}\right)\chi_{\bar{q}}(\beta\omega)\bar{A}_f}{\sum_f g_f\int\frac{d^3{\rm p}}{(2\pi)^3} ~ \omega^2_f \bar{f^q_0}\left(1-\bar{f^q_0}\right)\chi_{\bar{q}}(\beta\omega)}, \\ 
&&b_g=\frac{g_g\int\frac{d^3{\rm p}}{(2\pi)^3} ~ \omega_g f^g_0\left(1+f^g_0\right)\chi_g(\beta\omega)A_g}{g_g\int\frac{d^3{\rm p}}{(2\pi)^3} ~ \omega^2_g f^g_0\left(1+f^g_0\right)\chi_g(\beta\omega)}
.\end{eqnarray}
Substituting $A_f\rightarrow A_f^\prime$, $\bar{A}_f\rightarrow \bar{A}_f^\prime$ and $A_g\rightarrow A_g^\prime$ in eq. \eqref{iso.zeta1 (1)} and then simplifying, we get the bulk viscosity for a 
rotating QGP medium in the relaxation time approximation as
\begin{eqnarray}\label{iso.zeta (1)}
\nonumber\zeta &=& \frac{\beta}{18\pi^2}\sum_f g_f \int d{\rm p}~{\rm p}^2\left[\frac{{\rm p}^2}{\omega_f}-3\left(\frac{\partial P}{\partial \varepsilon}\right)\omega_f\right]^2\left[t_f f^q_0\left(1-f^q_0\right)\chi_q(\beta\omega)+t_{\bar{f}} \bar{f^q_0}\left(1-\bar{f^q_0}\right)\chi_{\bar{q}}(\beta\omega)\right] \\ && +\frac{\beta}{18\pi^2}g_g\int d{\rm p}~{\rm p}^2\left[\frac{{\rm p}^2}{\omega_g}-3\left(\frac{\partial P}{\partial \varepsilon}\right)\omega_g\right]^2 t_g f^g_0\left(1+f^g_0\right)\chi_g(\beta\omega)
~.\end{eqnarray}

\noindent{\bf\underline{Novel relaxation time approximation method:}} 

\noindent In the novel relaxation time approximation method, the collision integral takes a modified form to preserve the essential characteristics of the linearized Boltzmann collision operator. Consequently, the microscopic conservation laws can be preserved when attempting to describe relativistic gases using the momentum-dependent relaxation time or general matching conditions \cite{Rocha:PRL127'2021}. In the novel RTA method, relativistic Boltzmann transport equations for quarks, antiquarks and gluons are expressed as
\be\label{R.B.T.E.Q}
\nonumber p^\mu\partial_\mu f_q^\prime=-\frac{p_\nu u^\nu}{\tau_f}\left[\delta f_q-\frac{\left\langle\left(\omega_f/\tau_f\right)\delta f_q\right\rangle_0}{\left\langle\omega_f/\tau_f\right\rangle_0}+\frac{P_1^{(0)}\left\langle\left(\omega_f/\tau_f\right)P_1^{(0)}\delta f_q\right\rangle_0}{\left\langle\left(\omega_f/\tau_f\right)P_1^{(0)}P_1^{(0)}\right\rangle_0}\right. \\ \left.+\frac{p^{\left\langle\mu\right\rangle}\left\langle\left(\omega_f/\tau_f\right)p_{\left\langle\mu\right\rangle}\delta f_q\right\rangle_0}{(1/3)\left\langle\left(\omega_f/\tau_f\right)p_{\left\langle\mu\right\rangle}p^{\left\langle\mu\right\rangle}\right\rangle_0}\right]
,\ee
\be\label{R.B.T.E.A}
\nonumber p^\mu\partial_\mu \bar{f}_q^\prime=-\frac{p_\nu u^\nu}{\tau_{\bar{f}}}\left[\delta \bar{f}_q-\frac{\left\langle\left(\omega_f/\tau_{\bar{f}}\right)\delta \bar{f}_q\right\rangle_0}{\left\langle\omega_f/\tau_{\bar{f}}\right\rangle_0}+\frac{P_1^{(0)}\left\langle\left(\omega_f/\tau_{\bar{f}}\right)P_1^{(0)}\delta \bar{f}_q\right\rangle_0}{\left\langle\left(\omega_f/\tau_{\bar{f}}\right)P_1^{(0)}P_1^{(0)}\right\rangle_0}\right. \\ \left.+\frac{p^{\left\langle\mu\right\rangle}\left\langle\left(\omega_f/\tau_{\bar{f}}\right)p_{\left\langle\mu\right\rangle}\delta \bar{f}_q\right\rangle_0}{(1/3)\left\langle\left(\omega_f/\tau_{\bar{f}}\right)p_{\left\langle\mu\right\rangle}p^{\left\langle\mu\right\rangle}\right\rangle_0}\right]
,\ee
\be\label{R.B.T.E.G}
\nonumber p^\mu\partial_\mu f_g^\prime = -\frac{p_\nu u^\nu}{\tau_g}\left[\delta f_g-\frac{\left\langle\left(\omega_g/\tau_g\right)\delta f_g\right\rangle_0}{\left\langle\omega_g/\tau_g\right\rangle_0}+\frac{P_1^{(0)}\left\langle\left(\omega_g/\tau_g\right)P_1^{(0)}\delta f_g\right\rangle_0}{\left\langle\left(\omega_g/\tau_g\right)P_1^{(0)}P_1^{(0)}\right\rangle_0}\right. \\ \left.+\frac{p^{\left\langle\mu\right\rangle}\left\langle\left(\omega_g/\tau_g\right)p_{\left\langle\mu\right\rangle}\delta f_g\right\rangle_0}{(1/3)\left\langle\left(\omega_g/\tau_g\right)p_{\left\langle\mu\right\rangle}p^{\left\langle\mu\right\rangle}\right\rangle_0}\right]
,\ee
respectively. The novel relaxation times for quark (antiquark), $\tau_f$ ($\tau_{\bar{f}}$) and gluon, $\tau_g$ are momentum-dependent and take the following forms, 
\be\label{N.R.T.Q}
\tau_{f(\bar{f})}=\left(\beta\omega_f\right)^\xi t_{f(\bar{f})}, \\ 
\label{N.R.T.G}\tau_g=\left(\beta\omega_g\right)^\xi t_g 
,\ee
where $\xi$ is an arbitrary constant that controls the energy dependence of the relaxation times, while $t_{f(\bar{f})}$ and $t_g$ are the momentum-independent relaxation times, whose explicit forms are 
given in equations \eqref{R.T.Q} and \eqref{R.T.G}, respectively. The value of $\xi$ is different for different theories. For example, in QCD kinetic theories, $\xi=\frac{1}{2}$ \cite{Dusling:PRC81'2010}, while in scalar field theories, $\xi=1$ \cite{Calzetta:PRD37'1988}. In equations \eqref{R.B.T.E.Q}, \eqref{R.B.T.E.A} and \eqref{R.B.T.E.G}, $P_1^{(0)}$ and $p^{\left\langle\mu\right\rangle}$ are defined as
\be\label{E.1}
&&P_1^{(0)}=1-\frac{\left\langle\omega_f/\tau_f\right\rangle_0\omega_f}{\left\langle\omega_f^2/\tau_f\right\rangle_0} , \\ 
&&\label{E.2}p^{\left\langle\mu\right\rangle}=\Delta^{\mu\nu}p_\nu
,\ee
where $\Delta^{\mu\nu}=g^{\mu\nu}-u^\mu u^\nu$. The momentum integral of a quantity (say $A$) relative to the local equilibrium distribution function can be defined as
\be\label{E.3}
\left\langle A \right\rangle_0=\int\frac{d^3\rm{p}}{(2\pi)^3\omega_f}Af_q
~.\ee
Using equations \eqref{E.1}, \eqref{E.2}, \eqref{E.3} and integration by parts, we solve the relativistic Boltzmann transport equations \eqref{R.B.T.E.Q}, \eqref{R.B.T.E.A} and \eqref{R.B.T.E.G} to obtain the infinitesimal changes in the parton distribution functions as
\be
\delta f_q=-\frac{\tau_f p^\mu\partial_\mu f_q^\prime}{p_\nu u^\nu J}, \\ 
\delta \bar{f}_q=-\frac{\tau_{\bar{f}} p^\mu\partial_\mu \bar{f}_q^\prime}{p_\nu u^\nu \bar{J}}, \\ 
\delta f_g=-\frac{\tau_g p^\mu\partial_\mu f_g^\prime}{p_\nu u^\nu J_g}
,\ee
where the expressions for $J$, $\bar{J}$ and $J_g$ are given by
\be\label{Q.J}
J=\frac{\left(1-\frac{\omega_f\int d{\rm p}~\frac{{\rm p}^2}{\tau_f}f_q}{\int d{\rm p}~\frac{{\rm p}^2}{\tau_f}\omega_f f_q}\right)\int d{\rm p}~\frac{{\rm p}^2}{\tau_f}f_q\left(1-\frac{\omega_f\int d{\rm p}~\frac{{\rm p}^2}{\tau_f}f_q}{\int d{\rm p}~\frac{{\rm p}^2}{\tau_f}\omega_f f_q}\right)}{\int d{\rm p}~\frac{{\rm p}^2}{\tau_f}f_q\left(1-\frac{\omega_f\int d{\rm p}~\frac{{\rm p}^2}{\tau_f}f_q}{\int d{\rm p}~\frac{{\rm p}^2}{\tau_f}\omega_f f_q}\right)^2}+\frac{3{\rm p}\int d{\rm p}~\frac{{\rm p}^3}{\tau_f}f_q}{\int d{\rm p}~\frac{{\rm p}^4}{\tau_f}f_q}, \\ 
\label{A.J}\bar{J}=\frac{\left(1-\frac{\omega_f\int d{\rm p}~\frac{{\rm p}^2}{\tau_{\bar{f}}}\bar{f_q}}{\int d{\rm p}~\frac{{\rm p}^2}{\tau_{\bar{f}}}\omega_f \bar{f_q}}\right)\int d{\rm p}~\frac{{\rm p}^2}{\tau_{\bar{f}}}\bar{f_q}\left(1-\frac{\omega_f\int d{\rm p}~\frac{{\rm p}^2}{\tau_{\bar{f}}}\bar{f_q}}{\int d{\rm p}~\frac{{\rm p}^2}{\tau_{\bar{f}}}\omega_f \bar{f_q}}\right)}{\int d{\rm p}~\frac{{\rm p}^2}{\tau_{\bar{f}}}\bar{f_q}\left(1-\frac{\omega_f\int d{\rm p}~\frac{{\rm p}^2}{\tau_{\bar{f}}}\bar{f_q}}{\int d{\rm p}~\frac{{\rm p}^2}{\tau_{\bar{f}}}\omega_f \bar{f_q}}\right)^2}+\frac{3{\rm p}\int d{\rm p}~\frac{{\rm p}^3}{\tau_{\bar{f}}}\bar{f_q}}{\int d{\rm p}~\frac{{\rm p}^4}{\tau_{\bar{f}}}\bar{f_q}}, \\ 
\label{G.J}J_g=\frac{\left(1-\frac{\omega_g\int d{\rm p}~\frac{{\rm p}^2}{\tau_g}f_g}{\int d{\rm p}~\frac{{\rm p}^2}{\tau_g}\omega_g f_g}\right)\int d{\rm p}~\frac{{\rm p}^2}{\tau_g}f_g\left(1-\frac{\omega_g\int d{\rm p}~\frac{{\rm p}^2}{\tau_g}f_g}{\int d{\rm p}~\frac{{\rm p}^2}{\tau_g}\omega_g f_g}\right)}{\int d{\rm p}~\frac{{\rm p}^2}{\tau_g}f_g\left(1-\frac{\omega_g\int d{\rm p}~\frac{{\rm p}^2}{\tau_g}f_g}{\int d{\rm p}~\frac{{\rm p}^2}{\tau_g}\omega_g f_g}\right)^2}+\frac{3{\rm p}\int d{\rm p}~\frac{{\rm p}^3}{\tau_g}f_g}{\int d{\rm p}~\frac{{\rm p}^4}{\tau_g}f_g}
.\ee
Using equations \eqref{Q.D.F.}, \eqref{A.D.F.}, \eqref{G.D.F.}, \eqref{N.R.T.Q} and \eqref{N.R.T.G} in equations \eqref{Q.J}, \eqref{A.J} and \eqref{G.J}, the expressions for $J$, $\bar{J}$ and $J_g$ 
simplify to 
\be\label{Q.J (1)}
&&J=\frac{\left(1-\frac{\omega_f\int d{\rm p}~\frac{{\rm p}^2}{\omega_f^\xi}f^q_0}{\int d{\rm p}~\frac{{\rm p}^2}{\omega_f^{\xi-1}}f^q_0}\right)\int d{\rm p}~\frac{{\rm p}^2}{\omega_f^\xi}f^q_0\left(1-\frac{\omega_f\int d{\rm p}~\frac{{\rm p}^2}{\omega_f^\xi}f^q_0}{\int d{\rm p}~\frac{{\rm p}^2}{\omega_f^{\xi-1}}f^q_0}\right)}{\int d{\rm p}~\frac{{\rm p}^2}{\omega_f^\xi}f^q_0\left(1-\frac{\omega_f\int d{\rm p}~\frac{{\rm p}^2}{\omega_f^\xi}f^q_0}{\int d{\rm p}~\frac{{\rm p}^2}{\omega_f^{\xi-1}}f^q_0}\right)^2}+\frac{3{\rm p}\int d{\rm p}~\frac{{\rm p}^3}{\omega_f^\xi}f^q_0}{\int d{\rm p}~\frac{{\rm p}^4}{\omega_f^\xi}f^q_0}, \\ 
&&\label{A.J (1)}\bar{J}=\frac{\left(1-\frac{\omega_f\int d{\rm p}~\frac{{\rm p}^2}{\omega_f^\xi}\bar{f^q_0}}{\int d{\rm p}~\frac{{\rm p}^2}{\omega_f^{\xi-1}}\bar{f^q_0}}\right)\int d{\rm p}~\frac{{\rm p}^2}{\omega_f^\xi}\bar{f^q_0}\left(1-\frac{\omega_f\int d{\rm p}~\frac{{\rm p}^2}{\omega_f^\xi}\bar{f^q_0}}{\int d{\rm p}~\frac{{\rm p}^2}{\omega_f^{\xi-1}}\bar{f^q_0}}\right)}{\int d{\rm p}~\frac{{\rm p}^2}{\omega_f^\xi}\bar{f^q_0}\left(1-\frac{\omega_f\int d{\rm p}~\frac{{\rm p}^2}{\omega_f^\xi}\bar{f^q_0}}{\int d{\rm p}~\frac{{\rm p}^2}{\omega_f^{\xi-1}}\bar{f^q_0}}\right)^2}+\frac{3{\rm p}\int d{\rm p}~\frac{{\rm p}^3}{\omega_f^\xi}\bar{f^q_0}}{\int d{\rm p}~\frac{{\rm p}^4}{\omega_f^\xi}\bar{f^q_0}}, \\ 
&&\label{G.J (1)} J_g=\frac{\left(1-\frac{\omega_g\int d{\rm p}~\frac{{\rm p}^2}{\omega_g^\xi}f^g_0}{\int d{\rm p}~\frac{{\rm p}^2}{\omega_g^{\xi-1}}f^g_0}\right)\int d{\rm p}~\frac{{\rm p}^2}{\omega_g^\xi}f^g_0\left(1-\frac{\omega_g\int d{\rm p}~\frac{{\rm p}^2}{\omega_g^\xi}f^g_0}{\int d{\rm p}~\frac{{\rm p}^2}{\omega_g^{\xi-1}}f^g_0}\right)}{\int d{\rm p}~\frac{{\rm p}^2}{\omega_g^\xi}f^g_0\left(1-\frac{\omega_g\int d{\rm p}~\frac{{\rm p}^2}{\omega_g^\xi}f^g_0}{\int d{\rm p}~\frac{{\rm p}^2}{\omega_g^{\xi-1}}f^g_0}\right)^2}+\frac{3{\rm p}\int d{\rm p}~\frac{{\rm p}^3}{\omega_g^\xi}f^g_0}{\int d{\rm p}~\frac{{\rm p}^4}{\omega_g^\xi}f^g_0}
.\ee
Substituting the expressions for $\delta f_q$, $\delta \bar{f}_q$ and $\delta f_g$ into eq. \eqref{em1}, we get 
\be\label{em2}
\Delta T^{\mu\nu}=-\int\frac{d^3{\rm p}}{(2\pi)^3}\frac{p^\mu p^\nu}{p_\nu u^\nu} \left[\sum_f g_f\left(\frac{\tau_f p^\mu\partial_\mu f_q^\prime}{\omega_fJ}+\frac{\tau_{\bar{f}} p^\mu\partial_\mu \bar{f}_q^\prime}{\omega_f\bar{J}}\right)+g_g\frac{\tau_g p^\mu\partial_\mu f_g^\prime}{\omega_gJ_g}\right]
.\ee
The partial derivatives of the nonequilibrium quark, antiquark and gluon distribution functions appearing 
in the above equation are respectively calculated as
\begin{eqnarray}
\nonumber\partial_\mu f_q^\prime = \beta f^q_0\left(1-f^q_0\right)\chi_q(\beta\omega)\left[u_\alpha p^\alpha u_\mu\frac{DT}{T}+u_\alpha p^\alpha\frac{\nabla_\mu T}{T}-u_\mu p^\alpha Du_\alpha\right. \\ \left.-p^\alpha\nabla_\mu u_\alpha+T\partial_\mu\left(\frac{\mu}{T}\right)\right]+\beta f^q_0\sinh\left(\frac{\beta\omega}{2}\right)T\partial_\mu\left(\frac{\omega}{T}\right)
,\end{eqnarray}
\begin{eqnarray}
\nonumber\partial_\mu \bar{f}_q^\prime = \beta \bar{f^q_0}\left(1-\bar{f^q_0}\right)\chi_{\bar{q}}(\beta\omega)\left[u_\alpha p^\alpha u_\mu\frac{DT}{T}+u_\alpha p^\alpha\frac{\nabla_\mu T}{T}-u_\mu p^\alpha Du_\alpha\right. \\ \left.-p^\alpha\nabla_\mu u_\alpha-T\partial_\mu\left(\frac{\mu}{T}\right)\right]+\beta \bar{f^q_0}\sinh\left(\frac{\beta\omega}{2}\right)T\partial_\mu\left(\frac{\omega}{T}\right)
,\end{eqnarray}
\begin{eqnarray}
\nonumber\partial_\mu f_g^\prime = \beta f^g_0\left(1+f^g_0\right)\chi_g(\beta\omega)\left[u_\alpha p^\alpha u_\mu\frac{DT}{T}+u_\alpha p^\alpha\frac{\nabla_\mu T}{T}-u_\mu p^\alpha Du_\alpha-p^\alpha\nabla_\mu u_\alpha\right] \\ +2\beta f^g_0\sinh\left(\beta\omega\right)T\partial_\mu\left(\frac{\omega}{T}\right)
.\end{eqnarray}
Using the expressions for $\partial_\mu f_q^\prime$, $\partial_\mu \bar{f}_q^\prime$ and $\partial_\mu f_g^\prime$ in eq. \eqref{em2} and substituting $\frac{DT}{T}=-\left(\frac{\partial P}{\partial \varepsilon}\right)\nabla_\alpha u^\alpha$ and $Du_\alpha=\frac{\nabla_\alpha P}{\varepsilon+P}$ from the energy-momentum conservation, we have
\be\label{em3}
\nonumber\Delta T^{\mu\nu} &=& \sum_f g_f\int\frac{d^3{\rm p}}{(2\pi)^3}\frac{p^\mu p^\nu \beta}{\omega_f}\left[\left\lbrace\frac{\tau_f f^q_0\left(1-f^q_0\right)\chi_q(\beta\omega)}{J}\left(\omega_f\left(\frac{\partial P}{\partial \varepsilon}\right)\nabla_\alpha u^\alpha\right.\right.\right. \\ && \left.\left.\left.\nonumber+p^\alpha\left(\frac{\nabla_\alpha P}{\varepsilon+P}-\frac{\nabla_\alpha T}{T}\right)-\frac{Tp^\alpha}{\omega_f}\partial_\alpha\left(\frac{\mu}{T}\right)+\frac{p^\alpha p^\beta}{\omega_f}\nabla_\alpha u_\beta\right)\right.\right. \\ && \left.\left.\nonumber -\frac{f^q_0\tau_f}{\omega_f J}\sinh\left(\frac{\beta\omega}{2}\right)Tp^\alpha\partial_\alpha\left(\frac{\omega}{T}\right)\right\rbrace\right. \\ && \left.\nonumber+\left\lbrace\frac{\tau_{\bar{f}} \bar{f^q_0}\left(1-\bar{f^q_0}\right)\chi_{\bar{q}}(\beta\omega)}{\bar{J}}\left(\omega_f\left(\frac{\partial P}{\partial \varepsilon}\right)\nabla_\alpha u^\alpha\right.\right.\right. \\ && \left.\left.\left.\nonumber+p^\alpha\left(\frac{\nabla_\alpha P}{\varepsilon+P}-\frac{\nabla_\alpha T}{T}\right)+\frac{Tp^\alpha}{\omega_f}\partial_\alpha\left(\frac{\mu}{T}\right)+\frac{p^\alpha p^\beta}{\omega_f}\nabla_\alpha u_\beta\right)\right.\right. \\ && \left.\left.\nonumber -\frac{\bar{f^q_0}\tau_{\bar{f}}}{\omega_f \bar{J}}\sinh\left(\frac{\beta\omega}{2}\right)Tp^\alpha\partial_\alpha\left(\frac{\omega}{T}\right)\right\rbrace\right] \\ && \nonumber+g_g\int\frac{d^3{\rm p}}{(2\pi)^3}\frac{p^\mu p^\nu \beta}{\omega_g}\left[\frac{\tau_g f^g_0\left(1+f^g_0\right)\chi_g(\beta\omega)}{J_g}\left\lbrace\omega_g\left(\frac{\partial P}{\partial \varepsilon}\right)\nabla_\alpha u^\alpha\right.\right. \\ && \left.\left.+p^\alpha\left(\frac{\nabla_\alpha P}{\varepsilon+P}-\frac{\nabla_\alpha T}{T}\right)+\frac{p^\alpha p^\beta}{\omega_g}\nabla_\alpha u_\beta\right\rbrace-\frac{2f^g_0\tau_g}{\omega_g J_g}\sinh\left(\beta\omega\right)Tp^\alpha\partial_\alpha\left(\frac{\omega}{T}\right)\right]
.\ee
From eq. \eqref{em3}, the spatial components of $\Delta T^{\mu\nu}$ are expressed as
\be\label{em4}
\nonumber\Delta T^{ij} &=& \sum_f g_f\int\frac{d^3{\rm p}}{(2\pi)^3}\frac{p^i p^j \beta}{\omega_f}\left[\left\lbrace\frac{\tau_f f^q_0\left(1-f^q_0\right)\chi_q(\beta\omega)}{J}\left(\left(\omega_f\left(\frac{\partial P}{\partial \varepsilon}\right)-\frac{\rm p^2}{3\omega_f}\right)\partial_l u^l\right.\right.\right. \\ && \left.\left.\left.\nonumber -\frac{p^kp^l}{2\omega_f}W_{kl}-\frac{Tp^k}{\omega_f}\partial_k\left(\frac{\mu}{T}\right)+p^k\left(\frac{\partial_k P}{\varepsilon+P}-\frac{\partial_k T}{T} \right)\right)\right.\right. \\ && \left.\left.\nonumber -\frac{f^q_0\tau_f}{\omega_f J}\sinh\left(\frac{\beta\omega}{2}\right)Tp^k\partial_k\left(\frac{\omega}{T}\right)\right\rbrace\right. \\ && \left.\nonumber+\left\lbrace\frac{\tau_{\bar{f}} \bar{f^q_0}\left(1-\bar{f^q_0}\right)\chi_{\bar{q}}(\beta\omega)}{\bar{J}}\left(\left(\omega_f\left(\frac{\partial P}{\partial \varepsilon}\right)-\frac{\rm p^2}{3\omega_f}\right)\partial_l u^l\right.\right.\right. \\ && \left.\left.\left.\nonumber -\frac{p^kp^l}{2\omega_f}W_{kl}+\frac{Tp^k}{\omega_f}\partial_k\left(\frac{\mu}{T}\right)+p^k\left(\frac{\partial_k P}{\varepsilon+P}-\frac{\partial_k T}{T}\right)\right)\right.\right. \\ && \left.\left.\nonumber -\frac{\bar{f^q_0}\tau_{\bar{f}}}{\omega_f \bar{J}}\sinh\left(\frac{\beta\omega}{2}\right)Tp^k\partial_k\left(\frac{\omega}{T}\right)\right\rbrace\right] \\ && \nonumber+g_g\int\frac{d^3{\rm p}}{(2\pi)^3}\frac{p^i p^j \beta}{\omega_g}\left[\frac{\tau_g f^g_0\left(1+f^g_0\right)\chi_g(\beta\omega)}{J_g}\left\lbrace\left(\omega_g\left(\frac{\partial P}{\partial \varepsilon}\right)-\frac{\rm p^2}{3\omega_g}\right)\partial_l u^l\right.\right. \\ && \left.\left.-\frac{p^kp^l}{2\omega_g}W_{kl}+p^k\left(\frac{\partial_k P}{\varepsilon+P}-\frac{\partial_k T}{T}\right)\right\rbrace-\frac{2f^g_0\tau_g}{\omega_g J_g}\sinh\left(\beta\omega\right)Tp^k\partial_k\left(\frac{\omega}{T}\right)\right]
,\ee
where we have used $\partial_k u_l=-\frac{1}{2}W_{kl}-\frac{1}{3}\delta_{kl}\partial_j u^j$ and $W_{kl}=\partial_k u_l+\partial_l u_k-\frac{2}{3}\delta_{kl}\partial_j u^j$. Using equations \eqref{N.R.T.Q} 
and \eqref{N.R.T.G} into eq. \eqref{em4} and comparing with eq. \eqref{definition}, we get the shear viscosity for a rotating QGP medium in the novel relaxation time approximation as
\begin{eqnarray}\label{iso.eta}
\nonumber\eta &=& \frac{\beta^{1+\xi}}{30\pi^2}\sum_f g_f \int d{\rm p}~\frac{{\rm p}^6}{\omega_f^{2-\xi}}\left[\frac{t_f}{J}f^q_0\left(1-f^q_0\right)\chi_q(\beta\omega)+\frac{t_{\bar{f}}}{\bar{J}}\bar{f^q_0}\left(1-\bar{f^q_0}\right)\chi_{\bar{q}}(\beta\omega)\right] \\ && +\frac{\beta^{1+\xi}}{30\pi^2} g_g \int d{\rm p}~\frac{{\rm p}^6}{\omega_g^{2-\xi}}\frac{t_g}{J_g}f^g_0\left(1+f^g_0\right)\chi_g(\beta\omega)
~.\end{eqnarray}
Similarly, comparing equations \eqref{em4} and \eqref{definition}, the bulk viscosity is obtained as
\begin{eqnarray}\label{iso.zeta1}
\nonumber\zeta &=& \frac{1}{3}\sum_f g_f \int\frac{d^3{\rm p}}{(2\pi)^3}~\frac{{\rm p}^2}{\omega_f}\left[f^q_0\left(1-f^q_0\right)\chi_q(\beta\omega)A_f+\bar{f^q_0}\left(1-\bar{f^q_0}\right)\chi_{\bar{q}}(\beta\omega)\bar{A}_f\right] \\ && +\frac{1}{3}g_g \int\frac{d^3{\rm p}}{(2\pi)^3}~\frac{{\rm p}^2}{\omega_g}f^g_0\left(1+f^g_0\right)\chi_g(\beta\omega)A_g
~,\end{eqnarray}
where $A_f=\frac{\beta\tau_f}{J}\left[\frac{{\rm p}^2}{3\omega_f}-\left(\frac{\partial P}{\partial \varepsilon}\right)\omega_f\right]$, $\bar{A}_f=\frac{\beta\tau_{\bar{f}}}{\bar{J}}\left[\frac{{\rm p}^2}{3\omega_f}-\left(\frac{\partial P}{\partial \varepsilon}\right)\omega_f\right]$ and $A_g=\frac{\beta\tau_g}{J_g}\left[\frac{{\rm p}^2}{3\omega_g}-\left(\frac{\partial P}{\partial \varepsilon}\right)\omega_g\right]$. To satisfy the Landau-Lifshitz condition ($\Delta T^{00}=0$) in the local rest frame, the factors $A_f$, $\bar{A}_f$ and $A_g$ are modified as $A_f\rightarrow A_f^\prime=A_f-b_f\omega_f$, $\bar{A}_f\rightarrow \bar{A}_f^\prime=\bar{A}_f-\bar{b}_f\omega_f$ and $A_g\rightarrow A_g^\prime=A_g-b_g\omega_g$, where $b_f$, $\bar{b}_f$ and $b_g$ are arbitrary constants, which are associated with the particle number 
and energy conservations. From the ``00'' component of eq. \eqref{em3}, the Landau-Lifshitz conditions for $A_f$, $\bar{A}_f$ and $A_g$ are respectively expressed as
\begin{eqnarray}
&&\sum_f g_f\int\frac{d^3{\rm p}}{(2\pi)^3} ~ \omega_f f^q_0\left(1-f^q_0\right)\chi_q(\beta\omega)\left(A_f-b_f\omega_f\right)=0 \label{A_i} ~,~ \\ 
&&\sum_f g_f\int\frac{d^3{\rm p}}{(2\pi)^3} ~ \omega_f \bar{f^q_0}\left(1-\bar{f^q_0}\right)\chi_{\bar{q}}(\beta\omega)\left(\bar{A}_f-\bar{b}_f\omega_f\right)=0 \label{A_i.1} ~,~ \\ 
&&g_g\int\frac{d^3{\rm p}}{(2\pi)^3} ~ \omega_g f^g_0\left(1+f^g_0\right)\chi_g(\beta\omega)\left(A_g-b_g\omega_g\right)=0 \label{A_g}
~.\end{eqnarray}
One can calculate $b_f$, $\bar{b}_f$ and $b_g$ by solving equations \eqref{A_i}, \eqref{A_i.1} and \eqref{A_g}. Substituting $A_f\rightarrow A_f^\prime$, $\bar{A}_f\rightarrow \bar{A}_f^\prime$ and $A_g\rightarrow A_g^\prime$ in eq. \eqref{iso.zeta1}, and using equations \eqref{N.R.T.Q} 
and \eqref{N.R.T.G}, we get the bulk viscosity for a rotating QGP medium 
in the novel relaxation time approximation as
\begin{eqnarray}\label{iso.zeta}
\nonumber\zeta &=& \frac{\beta^{1+\xi}}{18\pi^2}\sum_f g_f \int d{\rm p}~{\rm p}^2\omega_f^\xi\left[\frac{{\rm p}^2}{\omega_f}-3\left(\frac{\partial P}{\partial \varepsilon}\right)\omega_f\right]^2\left[\frac{t_f}{J}f^q_0\left(1-f^q_0\right)\chi_q(\beta\omega)+\frac{t_{\bar{f}}}{\bar{J}}\bar{f^q_0}\left(1-\bar{f^q_0}\right)\chi_{\bar{q}}(\beta\omega)\right] \\ && +\frac{\beta^{1+\xi}}{18\pi^2}g_g\int d{\rm p}~{\rm p}^2\omega_g^\xi\left[\frac{{\rm p}^2}{\omega_g}-3\left(\frac{\partial P}{\partial \varepsilon}\right)\omega_g\right]^2\frac{t_g}{J_g}f^g_0\left(1+f^g_0\right)\chi_g(\beta\omega)
~.\end{eqnarray}

\section{Results and discussions}
This section presents the results for the shear viscosity and the bulk viscosity in a rotating QGP medium. We compare the results obtained using the RTA and the novel RTA methods in order to understand how different approximation methods affect these transport properties in a rotating medium. Our results also 
incorporate the thermal masses of partons within the quasiparticle description of the QGP medium. Interaction of partons with the surrounding medium mainly contributes to the generation of the thermal masses. For a QGP medium at finite chemical potential, the thermal masses (squared) of a quark with flavor $f$ and the gluon, 
up to one-loop, are given \cite{Braaten:PRD45'1992,Peshier:PRD66'2002,Blaizot:PRD72'2005} by
\be\label{Q.P.M.(Quark mass)}
&& m_{fT}^2=\frac{g^2T^2}{6}\left(1+\frac{\mu_f^2}{\pi^2T^2}\right), \\
&&\label{Q.P.M.(Gluon mass)}m_{gT}^2=\frac{g^2T^2}{6}\left(N_c+\frac{N_f}{2}+\frac{3}{2\pi^2T^2}\sum_f\mu_f^2\right)
,\ee
respectively. Here, $g$ represents the one-loop running coupling, which is expressed \cite{Kapusta:BOOK'2006} as
\begin{eqnarray}
g^2=4\pi\alpha_s=\frac{48\pi^2}{\left(11N_c-2N_f\right)\ln\left({\Lambda^2}/{\Lambda_{\rm\overline{MS}}^2}\right)}
~,\end{eqnarray}
where the modiﬁed minimal subtraction (MS) renormalization scale $\Lambda_{\rm\overline{MS}}=0.176$ GeV, the renormalization scale $\Lambda=2\pi\sqrt{T^2+\mu_f^2/\pi^2}$ for electrically charged particles (quarks and antiquarks) and $\Lambda=2 \pi T$ for gluons. In this work, the value of the chemical potential is 
fixed at 0.06 GeV. 

\begin{figure}[]
\begin{center}
\begin{tabular}{c c}
\includegraphics[width=7.4cm]{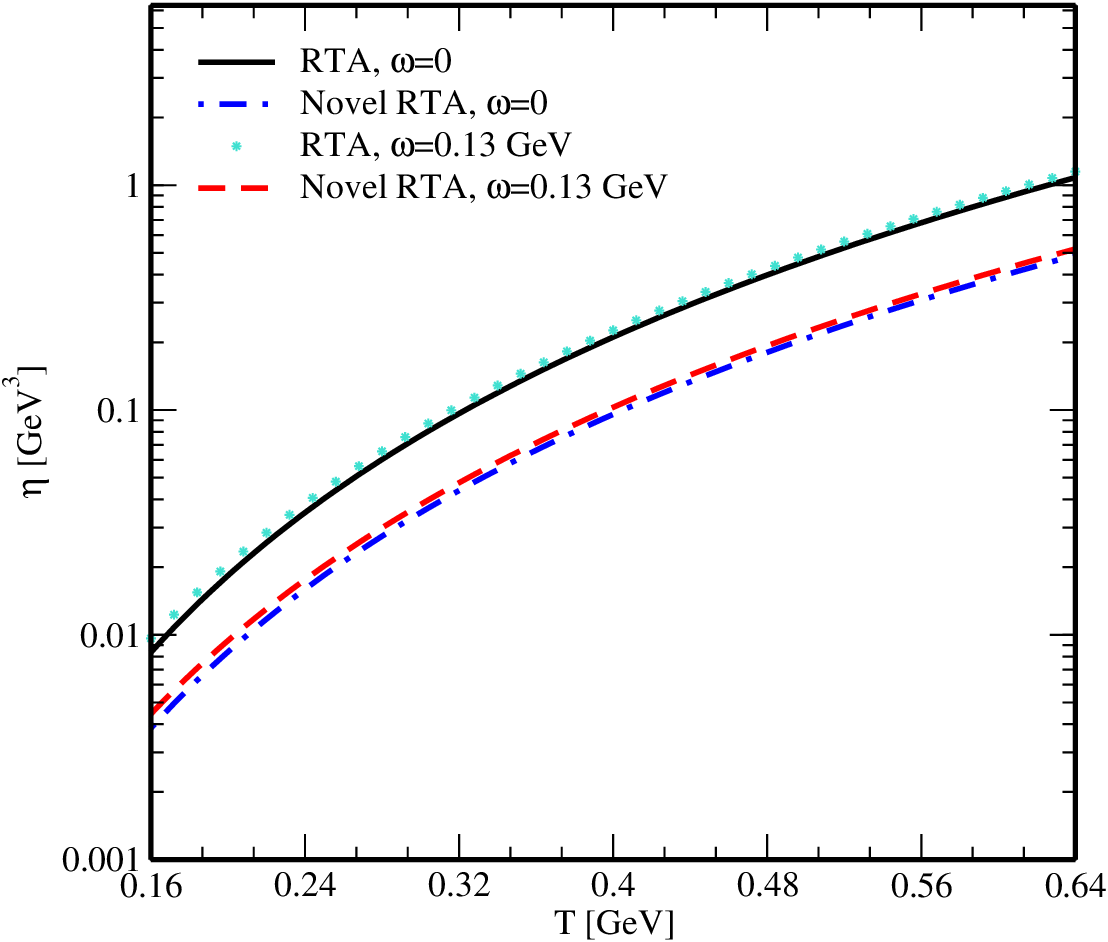}&
\hspace{0.73 cm}
\includegraphics[width=7.4cm]{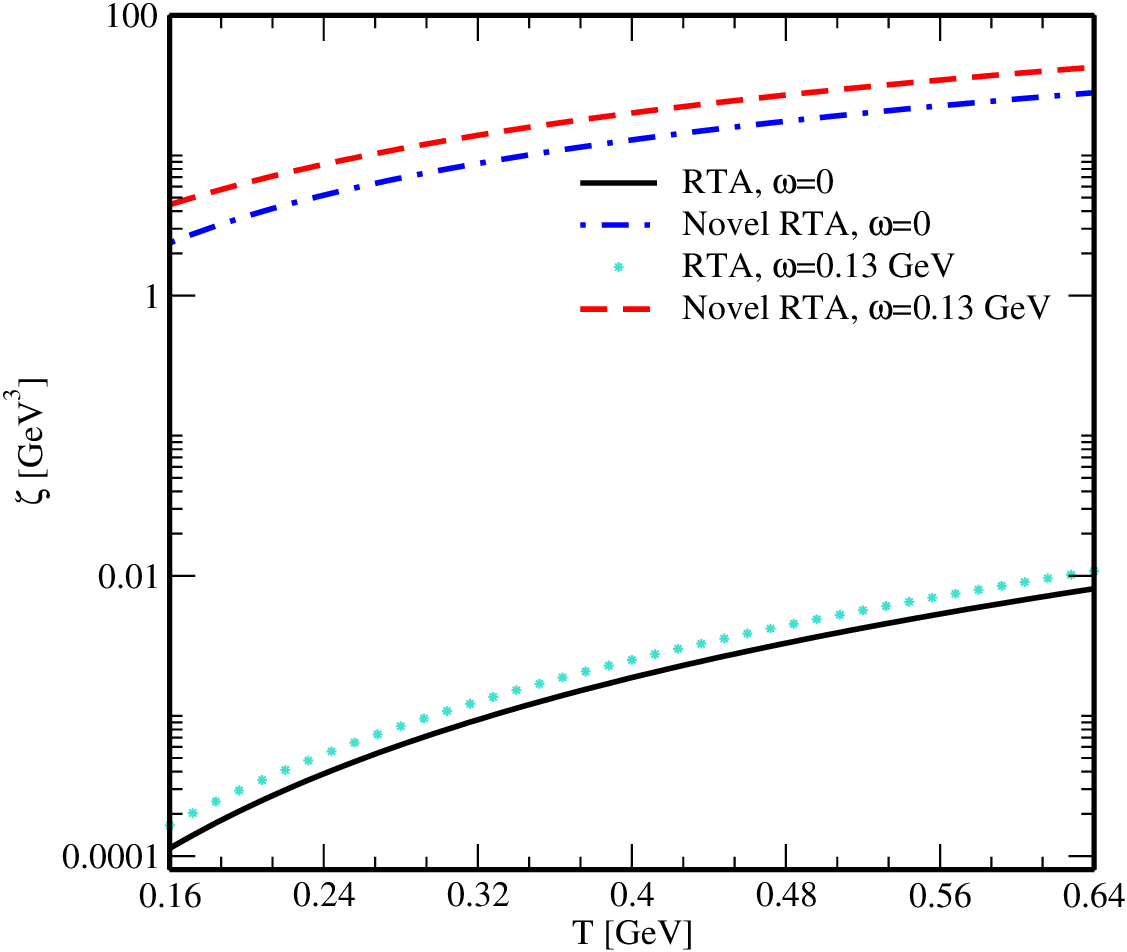} \\
\hspace{8mm}a & \hspace{18.1mm}b
\end{tabular}
\caption{Variation of (a) shear viscosity and (b) bulk viscosity as a function of temperature for different angular velocities, comparing the RTA and the novel RTA methods.}\label{Fig.1}
\end{center}
\end{figure}

Figure \ref{Fig.1} presents the variations of shear ($\eta$) and bulk ($\zeta$) viscosities of the QGP medium as a function of temperature for different values of angular velocity in the RTA method as well as 
in the novel RTA method. It is observed that the introduction of angular velocity enhances both $\eta$ (figure \ref{Fig.1}a) and $\zeta$ (figure \ref{Fig.1}b) over the considered temperature range. The increase in both shear and bulk viscosities at finite angular velocity confirms that the rotational feature of the medium amplifies the momentum transfer and the fluctuations in local pressure. Our study also finds that the influence of rotation on $\eta$ and $\zeta$ is more pronounced at low temperatures than at high temperatures. Within the novel RTA method, we observe a decrease in $\eta$ and an increase in $\zeta$ compared to the commonly used RTA method. This discrepancy arises primarily from differences in the relaxation times and collision integrals between the two approaches. Further, from figure \ref{Fig.1}, it is inferred that, within the standard RTA method, the shear viscosity remains nearly two orders of magnitude larger than the bulk viscosity for both rotating and nonrotating cases. As compared to the shear viscosity, the bulk viscosity is immensely influenced by the novel RTA method, exhibiting a substantial increase in magnitude. As a result, it is difficult to deform the system at a fixed volume as compared to the change in the volume of the system at a fixed shape, irrespective of whether the system is rotating or stationary. 

\section{Observable effects associated with the viscous properties}
This section explores some of the observable effects based on the shear and bulk viscous coefficients in a rotating QGP medium. Particularly, subsection 4.1 studies the impact of rotation on the flow 
characteristic, while subsection 4.2 discusses the impact of rotation on the specific shear and specific bulk viscosities. Finally, subsection 4.3 explores the Bjorken expansion of the rotating matter. 

\subsection{Flow characteristic}
Through the Reynolds number (Re), it is possible to understand the flow characteristic of the matter. The Reynolds number is linked to the kinematic viscosity (${\eta}/{\rho}$) as
\begin{equation}\label{Rl}
{\rm Re}=\frac{Lv}{\eta/\rho}
~,\end{equation}
where $L$, $v$ and $\rho$ represent the characteristic length of the system, the velocity of the flow and the mass density, respectively. The mass density can be determined using the number densities of quarks, 
antiquarks and gluons as
\begin{equation}\label{M.D.}
\rho=\sum_f m_f\left(n_f+\bar{n}_f\right)+m_gn_g
~,\end{equation}
where $m_f$ and $m_g$ are the thermal masses of charged particles (quarks or antiquarks) with flavor $f$ and gluons, respectively. At finite rotation, the mass density is calculated as
\begin{eqnarray}\label{M.D.}
\rho=\frac{1}{2\pi^2}\sum_f m_f g_f\int d{\rm p}~{\rm p}^2\left[f^q_0\chi_q(\beta\omega)+\bar{f^q_0}\chi_{\bar{q}}(\beta\omega)\right]+\frac{1}{2\pi^2}m_g g_g\int d{\rm p}~{\rm p}^2f^g_0\chi_g(\beta\omega)
.\end{eqnarray}
By setting $L=1$ fm and $v\simeq 1$, and using the values of $\eta$ from eq. \eqref{iso.eta} and $\rho$ from eq. \eqref{M.D.}, we have calculated the Reynolds number for a rotating QGP medium. The magnitude 
of the Reynolds number sheds light on the fluidity of the system. If the Reynolds number is much greater than unity, the nature of the flow is turbulent, whereas if the Reynolds number is small, the medium becomes a viscous system with laminar flow. Like other extreme conditions, rapid rotation could noticeably 
affect the Reynolds number, potentially altering the flow characteristic in such a regime. The present work explores the influence of rotation on the flow characteristic of the medium. 

\begin{figure}[]
\begin{center}
\begin{tabular}{c c}
\includegraphics[width=7.4cm]{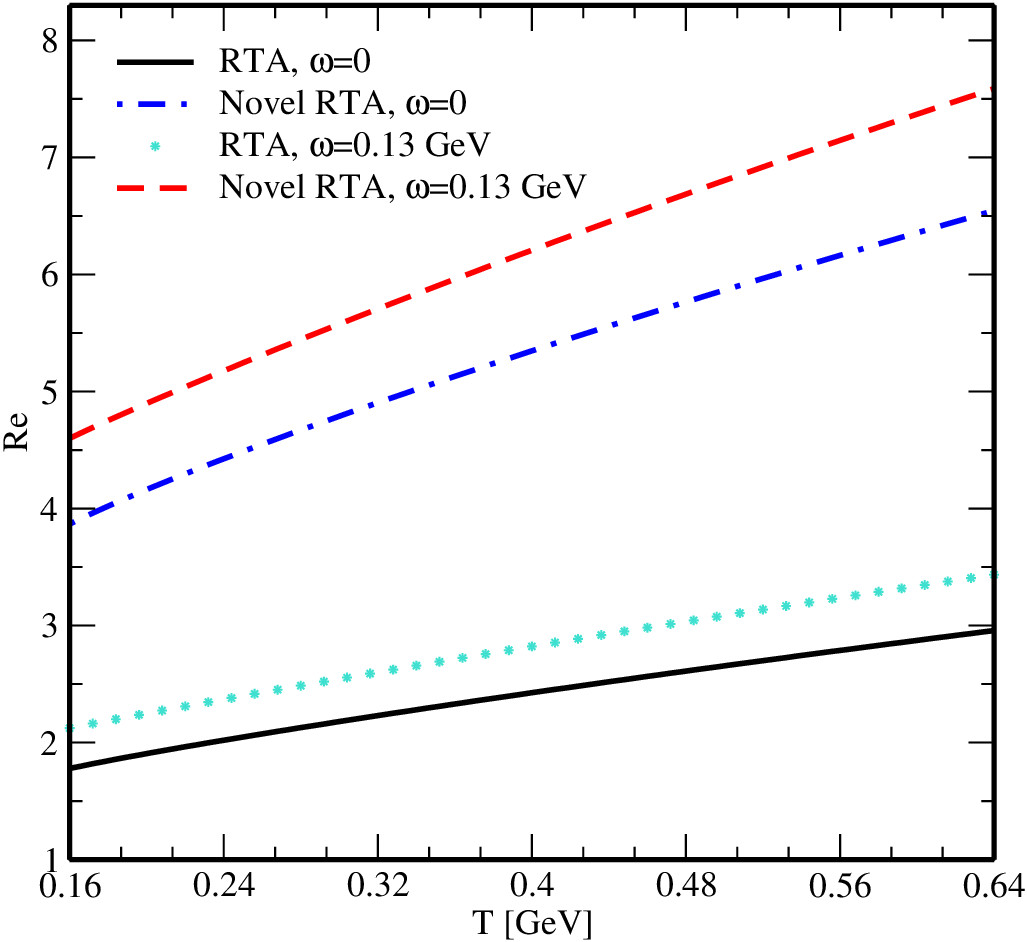}&
\hspace{0.73 cm}
\includegraphics[width=7.4cm]{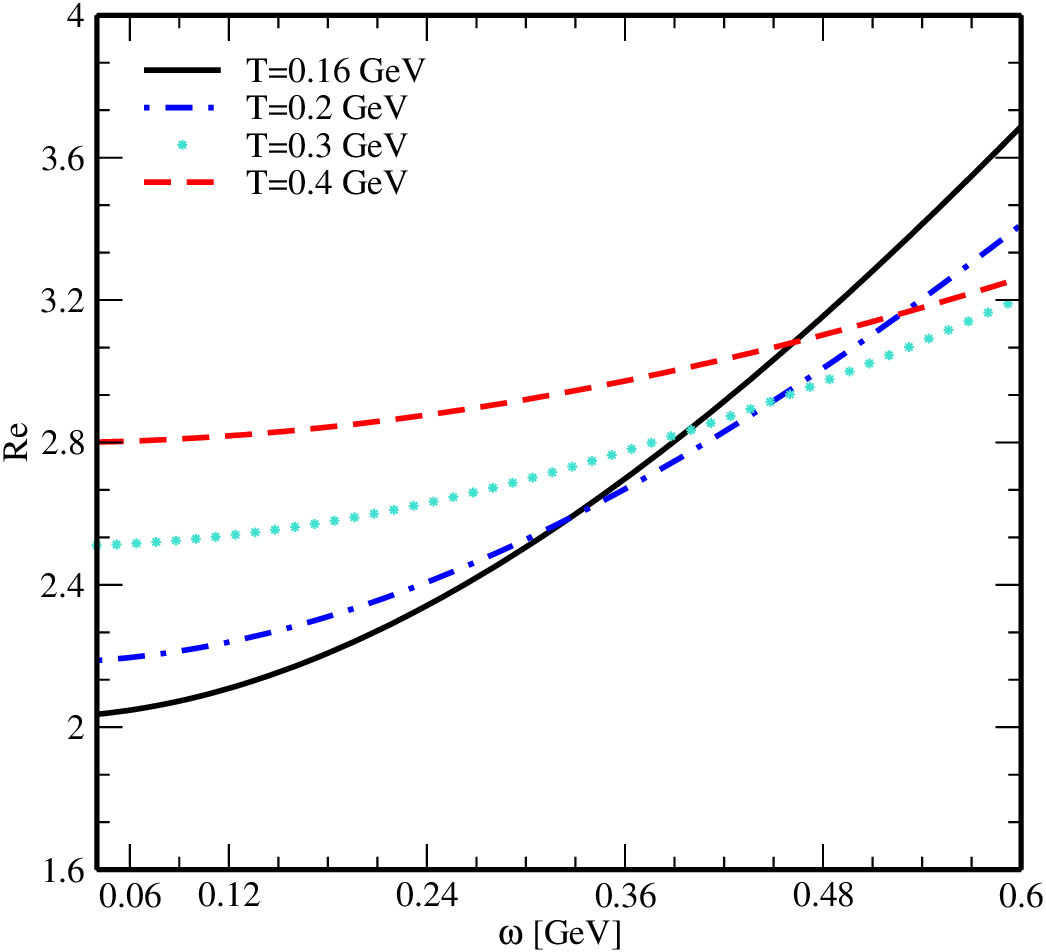} \\
\hspace{3mm}a & \hspace{13mm}b
\end{tabular}
\caption{Variation of the Reynolds number: (a) as a function of temperature for different angular velocities, comparing the RTA and the novel RTA methods, and (b) as a function of angular velocity for 
different temperatures within the RTA approach.}\label{Fig.2}
\end{center}
\end{figure}

\begin{figure}[]
\begin{center}
\begin{tabular}{c c}
\includegraphics[width=7.4cm]{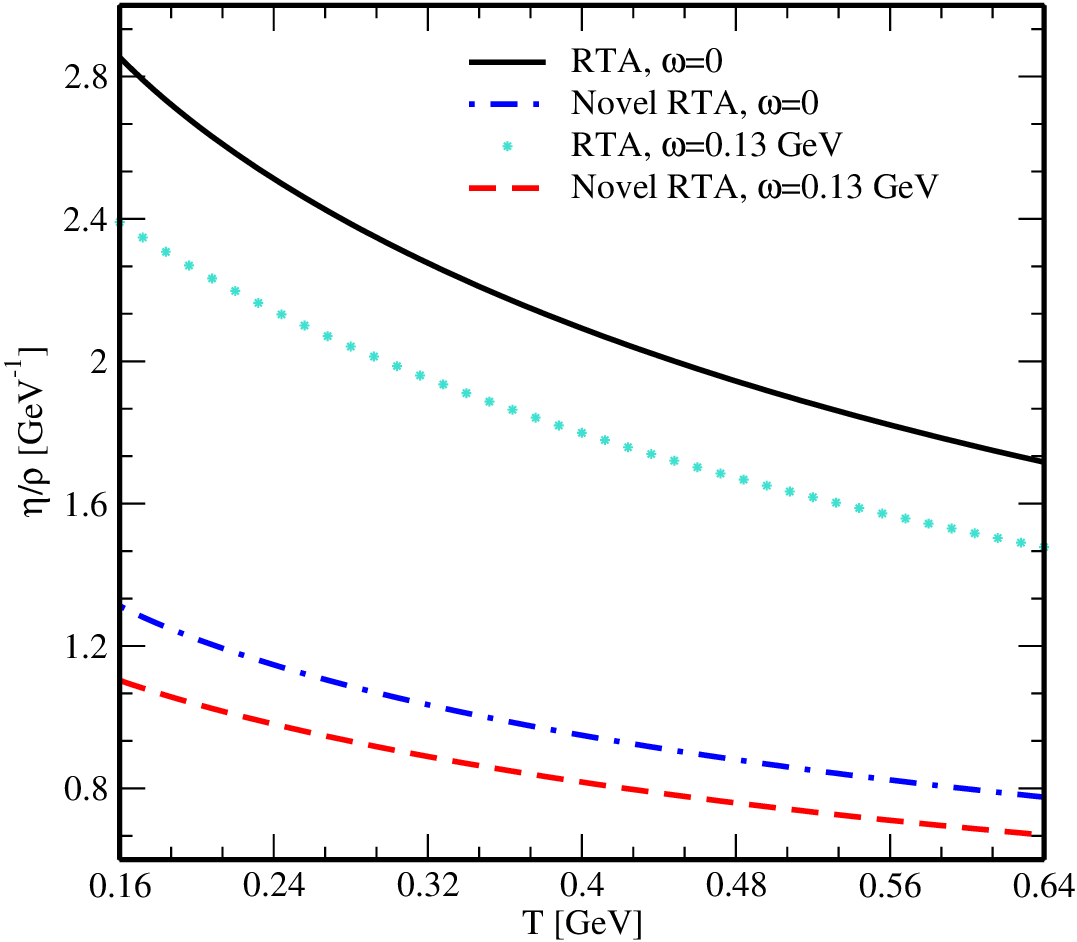}&
\hspace{0.73 cm}
\includegraphics[width=7.4cm]{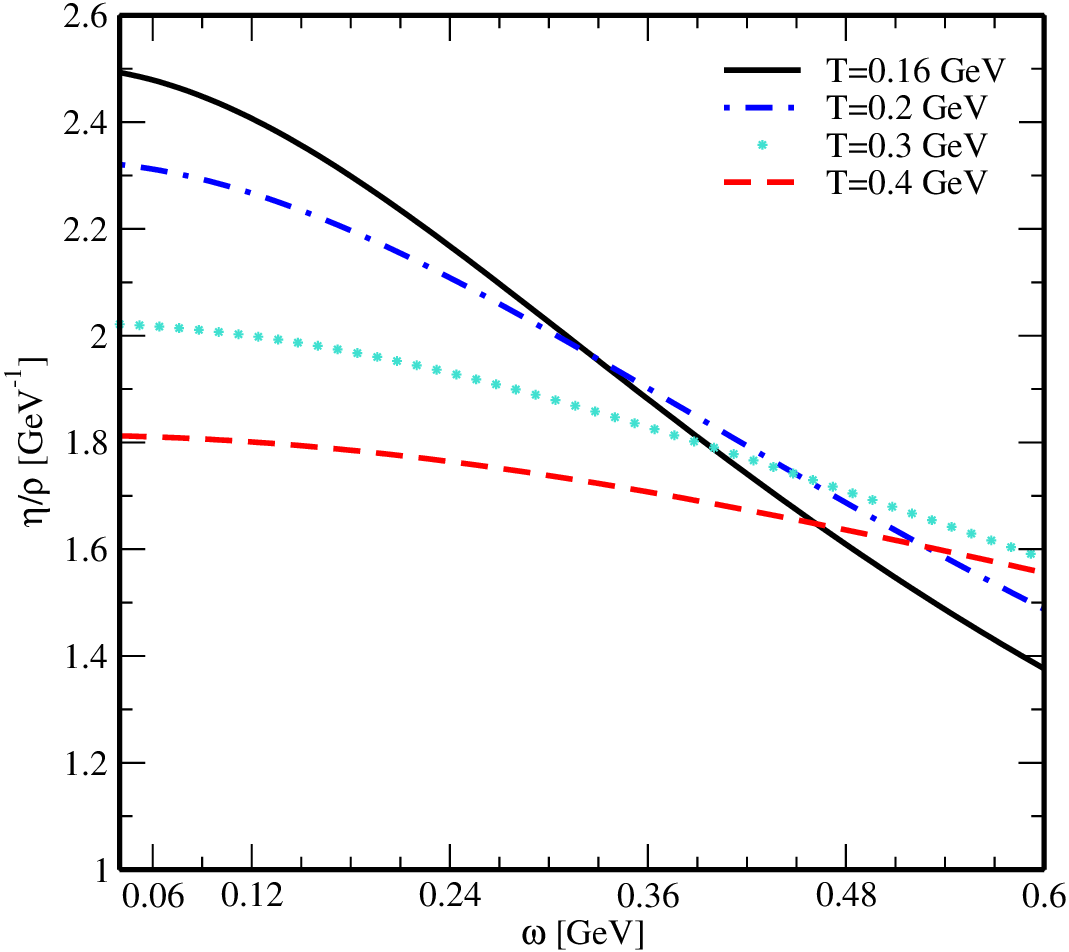} \\
\hspace{4.86mm}a & \hspace{16.1mm}b
\end{tabular}
\caption{Variation of kinematic viscosity: (a) as a function of temperature for different angular velocities, comparing the RTA and the novel RTA methods, and (b) as a function of angular velocity for 
different temperatures within the RTA approach.}\label{Fig.3}
\end{center}
\end{figure}

Figure \ref{Fig.2}a shows the variation of the Reynolds number as a function of temperature for different angular velocities using both the standard RTA and the novel RTA methods. Our results indicate an 
enhancement in the Reynolds number for a rapidly rotating QGP medium compared to the stationary case (figure \ref{Fig.2}a), which is consistent with the lower values of the kinematic viscosity observed at higher angular velocities (figure \ref{Fig.3}). In all cases, the novel RTA method predicts larger values of the Reynolds number than the standard RTA method, in accordance with the lower values of the kinematic viscosity observed in the novel RTA method (figure \ref{Fig.3}a). Figure \ref{Fig.2}b displays that, although the Reynolds number increases with angular velocity for all temperatures, this rise is not uniform, instead, it is significantly more pronounced at low temperatures than at high temperatures. This feature is mainly 
attributed to the nonuniform behavior of the kinematic viscosity (figure \ref{Fig.3}b). These findings 
suggest that the quark-gluon plasma medium becomes less viscous and retains its laminar flow characteristic at finite angular velocity. 

\subsection{Specific shear viscosity and specific bulk viscosity}
\begin{figure}[]
\begin{center}
\begin{tabular}{c c}
\includegraphics[width=7.4cm]{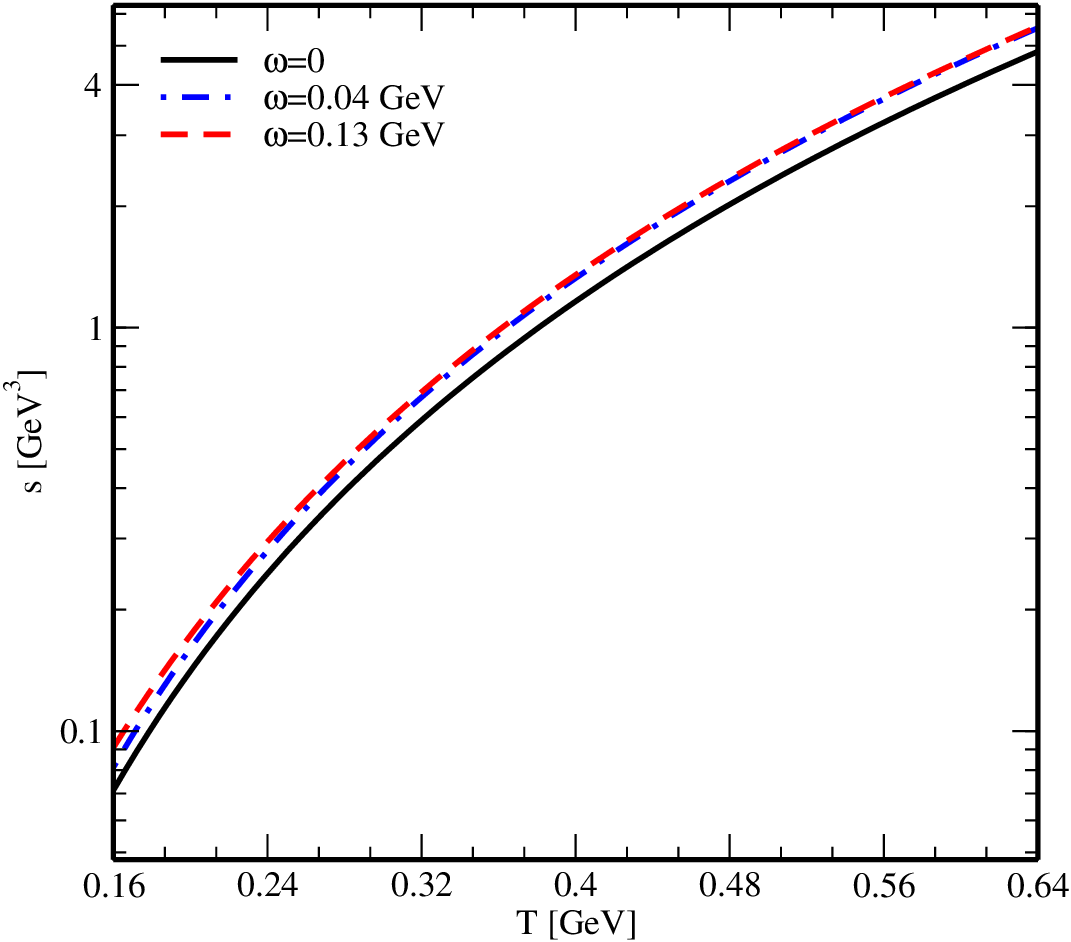}&
\hspace{0.73 cm}
\includegraphics[width=7.4cm]{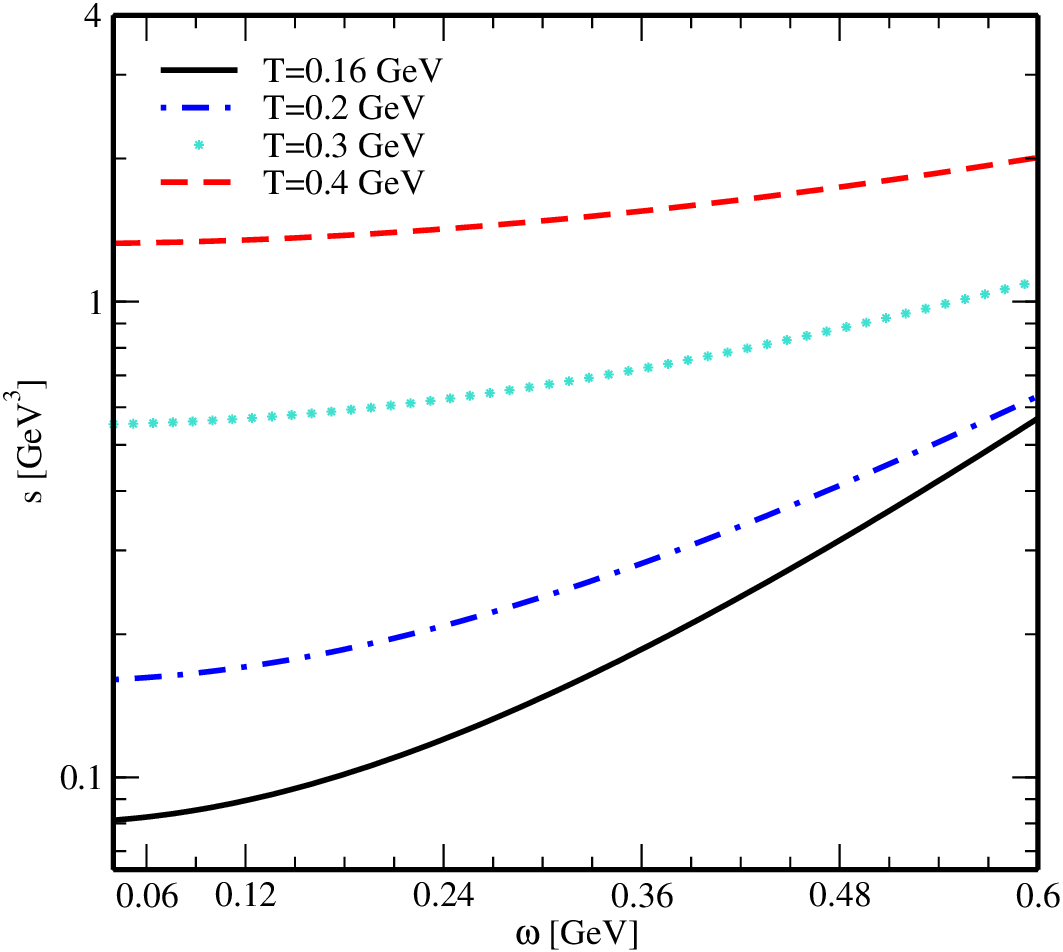} \\
\hspace{4.86mm}a & \hspace{16.1mm}b
\end{tabular}
\caption{Variation of entropy density: (a) as a function of temperature for different angular velocities and (b) as a function of angular velocity for different temperatures.}\label{Fig.4}
\end{center}
\end{figure}

In a medium, the specific shear ($\eta/s$) and specific bulk ($\zeta/s$) viscosities are essential in discerning the perfect fluid behavior and the conformal symmetry of the medium. These ratios for a 
rapidly rotating QGP medium mostly depend on how the properties of shear viscosity, bulk viscosity and entropy 
density get affected by the finite temperature and finite angular velocity. The entropy density ($s$) can be determined from its relation with the energy-momentum tensor and the baryon density ($n_B$) as
\begin{eqnarray}\label{E.D.}
s=\frac{u_\mu T^{\mu\nu}u_\nu-\sum_{f}\mu_f n_B-\Delta_{\mu\nu}T^{\mu\nu}/3}{T}
~.\end{eqnarray}
The baryon density for a rotating QGP medium is calculated as
\begin{eqnarray}
n_B &=& \frac{1}{2\pi^2}\sum_f g_f\int d{\rm p}~{\rm p}^2\left[f^q_0\chi_q(\beta\omega)-\bar{f^q_0}\chi_{\bar{q}}(\beta\omega)\right]
.\end{eqnarray}
The results for the entropy density at finite rotation and its comparison with the stationary case are shown in figure \ref{Fig.4}. The results indicate that, with the increase of angular velocity of the medium, the entropy density increases. Moreover, as the medium gets hotter, the deviation of the entropy density of stationary medium from that of rotating medium grows, suggesting that a faster rotation 
leads to a larger number of microstates. Thus, rotation makes the medium more disordered, whereas the disorderliness is comparatively small for a stationary medium. Additionally, it is found from figure \ref{Fig.4}a that the distinction between the blue ($\omega=0.04$ GeV) and red ($\omega=0.13$ GeV) lines is only apparent at low temperatures and these lines seem to coincide at high temperatures. This can be comprehended from the $\omega$-dependent factors appearing in the parton distribution functions. The shift of the $\omega$-dependent factor containing sine hyperbolic functions is small when changing the value of $\omega$ from 0.04 GeV to 0.13 GeV, thus contributing to a marginal shift of the entropy density. With the 
rising temperature, this shift diminishes further, indicating an almost overlap between $\omega=0.04$ GeV line and $\omega=0.13$ GeV line. Figure \ref{Fig.4}b explains that, within the rotating medium, the entropy density is observed to increase with the increase of angular velocity. Further, the trend of variation of entropy density with the angular velocity is not same for all temperatures, rather the increase of entropy density with angular velocity becomes slower for higher temperatures. This nonuniform variation can be understood from the fact that, for a rotating medium with low angular velocity, the temperature acts as the dominant energy 
scale, thus angular velocity leaves deficient impact on the phase-space distribution functions of partons, unlike the adequate impact in the high angular velocity regime. Consequently, a slower increase of entropy density with angular velocity is observed at higher temperatures. Studying the variations of entropy density, shear viscosity, and bulk viscosity at finite angular velocity helps to understand the ratios $\eta/s$ and $\zeta/s$ in this regime. 

\begin{figure}[]
\begin{center}
\begin{tabular}{c c}
\includegraphics[width=7.4cm]{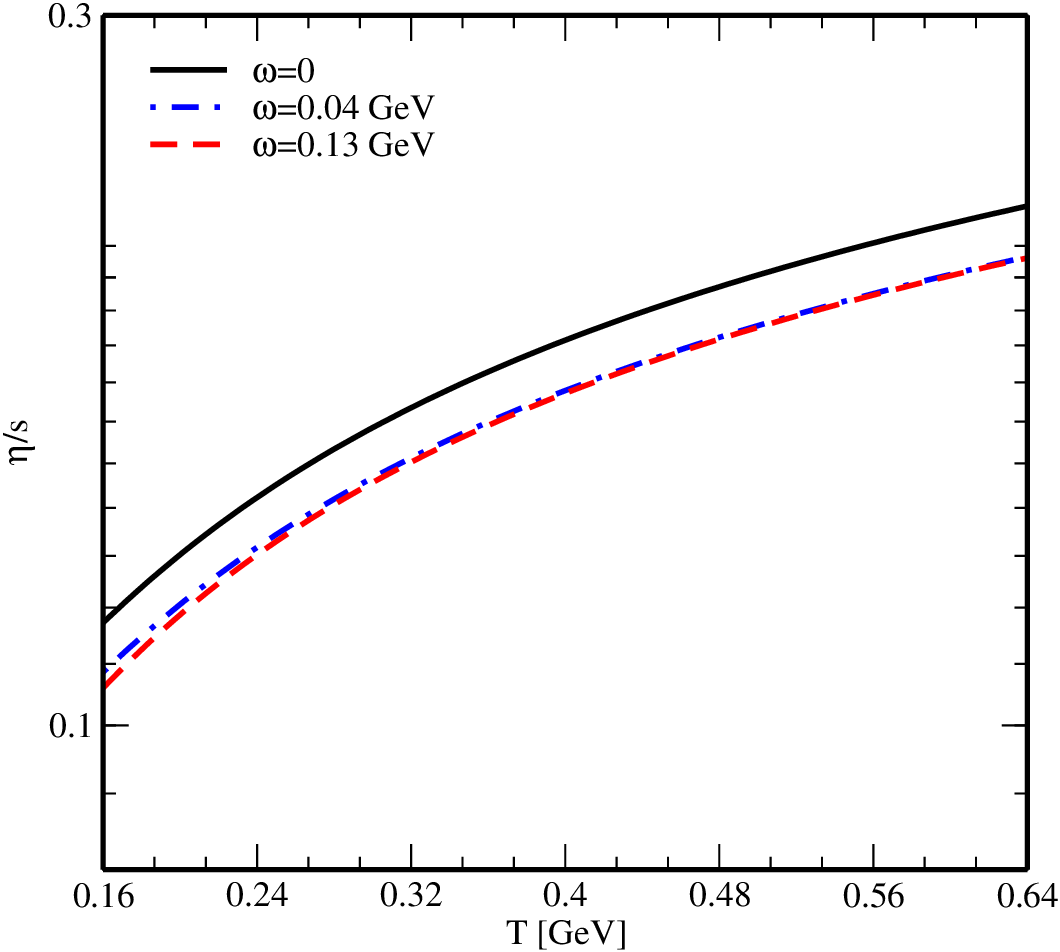}&
\hspace{0.73 cm}
\includegraphics[width=7.4cm]{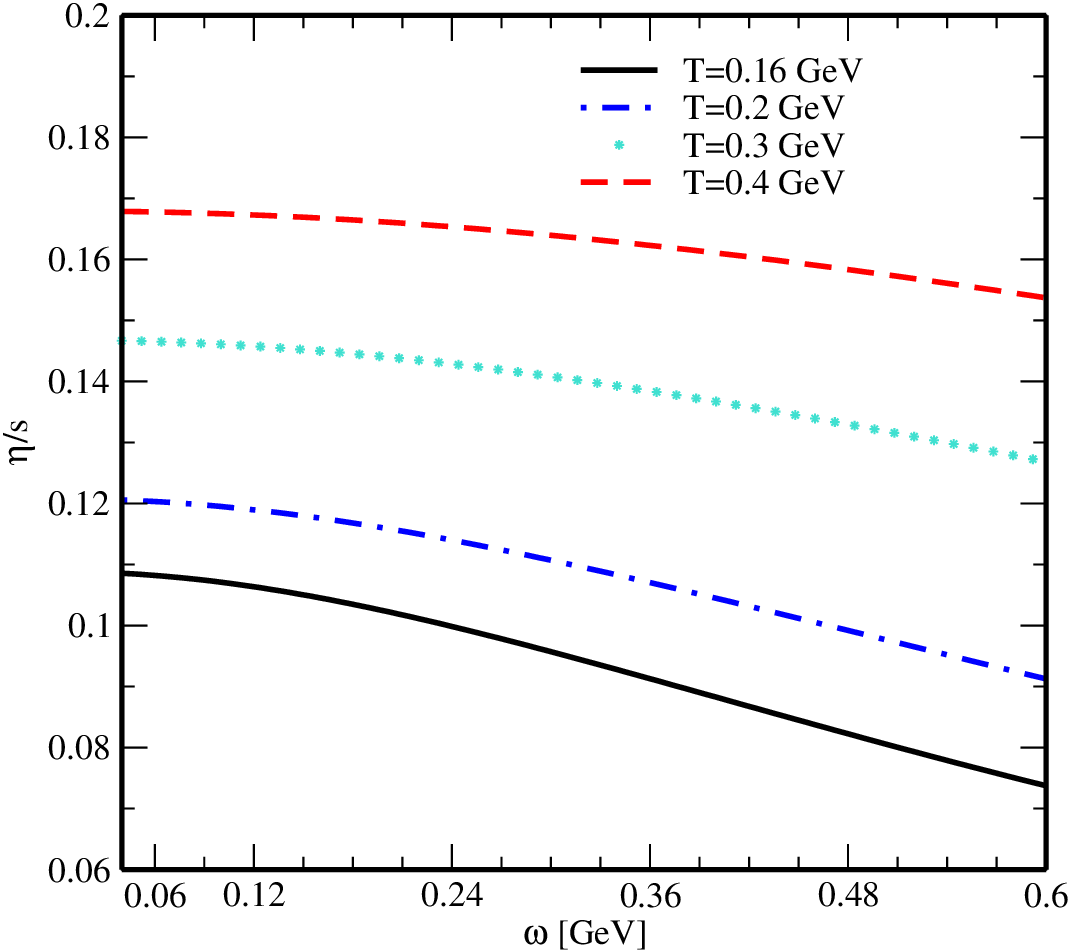} \\
\hspace{4.86mm}a & \hspace{16.1mm}b
\end{tabular}
\caption{Variation of the specific shear viscosity: (a) as a function of temperature for different angular velocities and (b) as a function of angular velocity for different temperatures.}\label{Fig.5}
\end{center}
\end{figure}

\begin{figure}[]
\begin{center}
\begin{tabular}{c c}
\includegraphics[width=7.4cm]{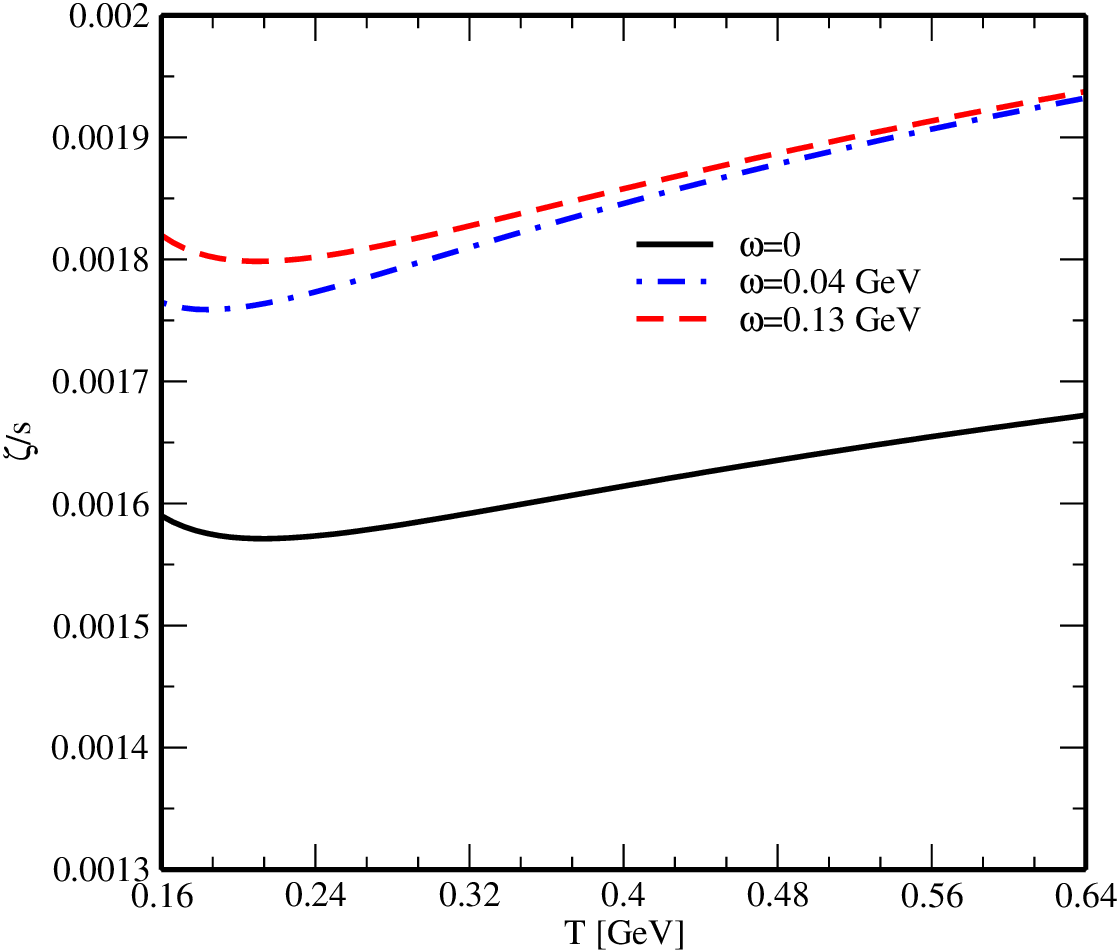}&
\hspace{0.73 cm}
\includegraphics[width=7.4cm]{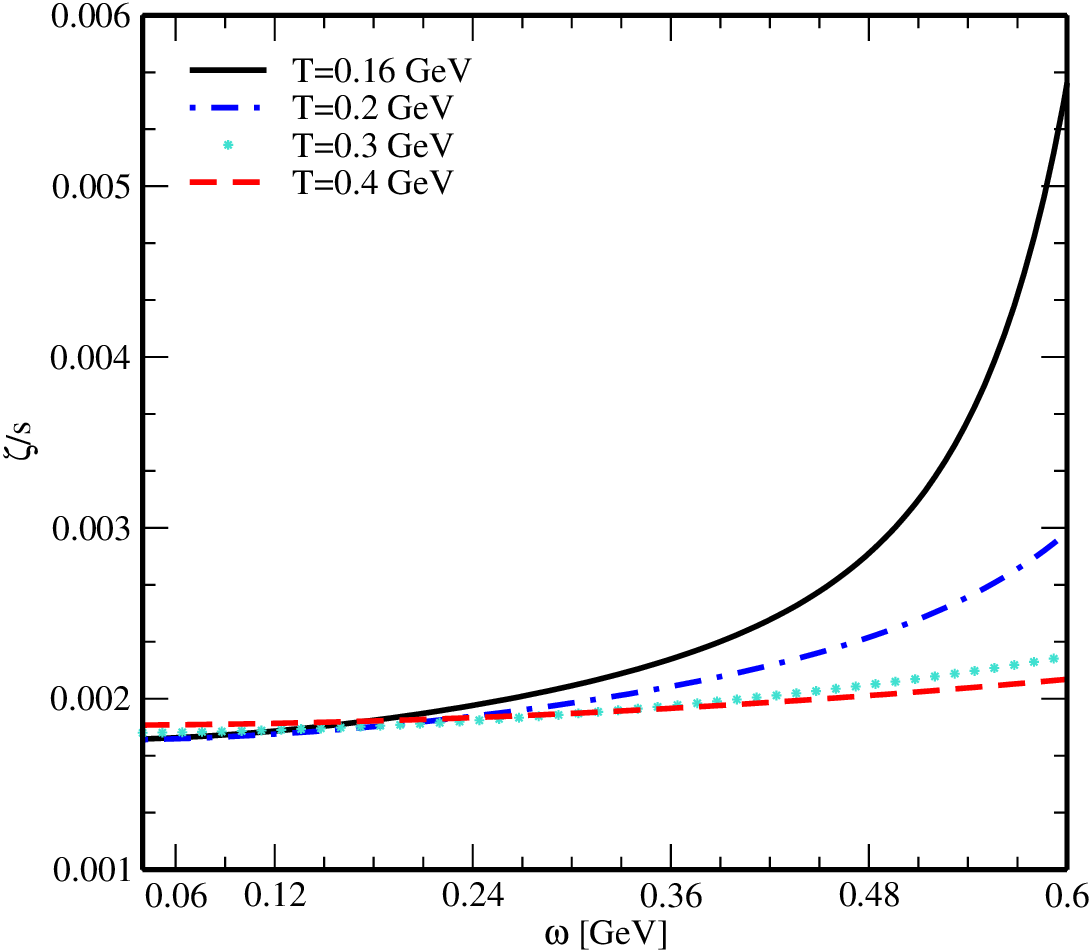} \\
\hspace{8mm}a & \hspace{17.1mm}b
\end{tabular}
\caption{Variation of the specific bulk viscosity: (a) as a function of temperature for different angular velocities and (b) as a function of angular velocity for different temperatures.}\label{Fig.6}
\end{center}
\end{figure}

Figures \ref{Fig.5} and \ref{Fig.6} show the variations of $\eta/s$ and $\zeta/s$ with temperature and angular velocity, respectively. A decreasing trend of $\eta/s$ with angular velocity is observed, in contrast to its rising trend with temperature (figure \ref{Fig.5}). At low temperatures, the effect of rotation on $\eta/s$ is more pronounced. On the other hand, as the temperature increases, the influence of rotation on $\eta/s$ gets suppressed. This suppression can be understood from the fact that, at high temperatures, the energy scale associated with the temperature dominates over the energy scale related to the angular velocity. Furthermore, the rotation of the QGP medium pushes the ratio $\eta/s$ closer to the conjectured lower bound $1/(4\pi)$, indicating a nearly perfect fluid characteristic of the medium. Generally, in a conformally symmetric medium, the ratio $\zeta/s$ approaches zero. However, our results show that the rotation enhances the value of $\zeta/s$ compared to the stationary case (figure \ref{Fig.6}), with this effect being more pronounced at 
low temperatures than at high temperatures. Consequently, the medium deviates away from conformal symmetry as angular velocity increases. 

\subsection{Bjorken expansion of rotating matter}
For a rotating medium produced in ultrarelativistic heavy-ion collisions, there exist plausible effects of angular velocity on both its static and dynamic properties. The rapid rotation exists for a longer time 
due to the conservation of the total angular momentum. This motivates us to study the effect of rotation 
on the hydrodynamic evolution of matter produced in ultrarelativistic heavy-ion collisions. 

\begin{figure}[]
\begin{center}
\begin{tabular}{c c}
\includegraphics[width=7.4cm]{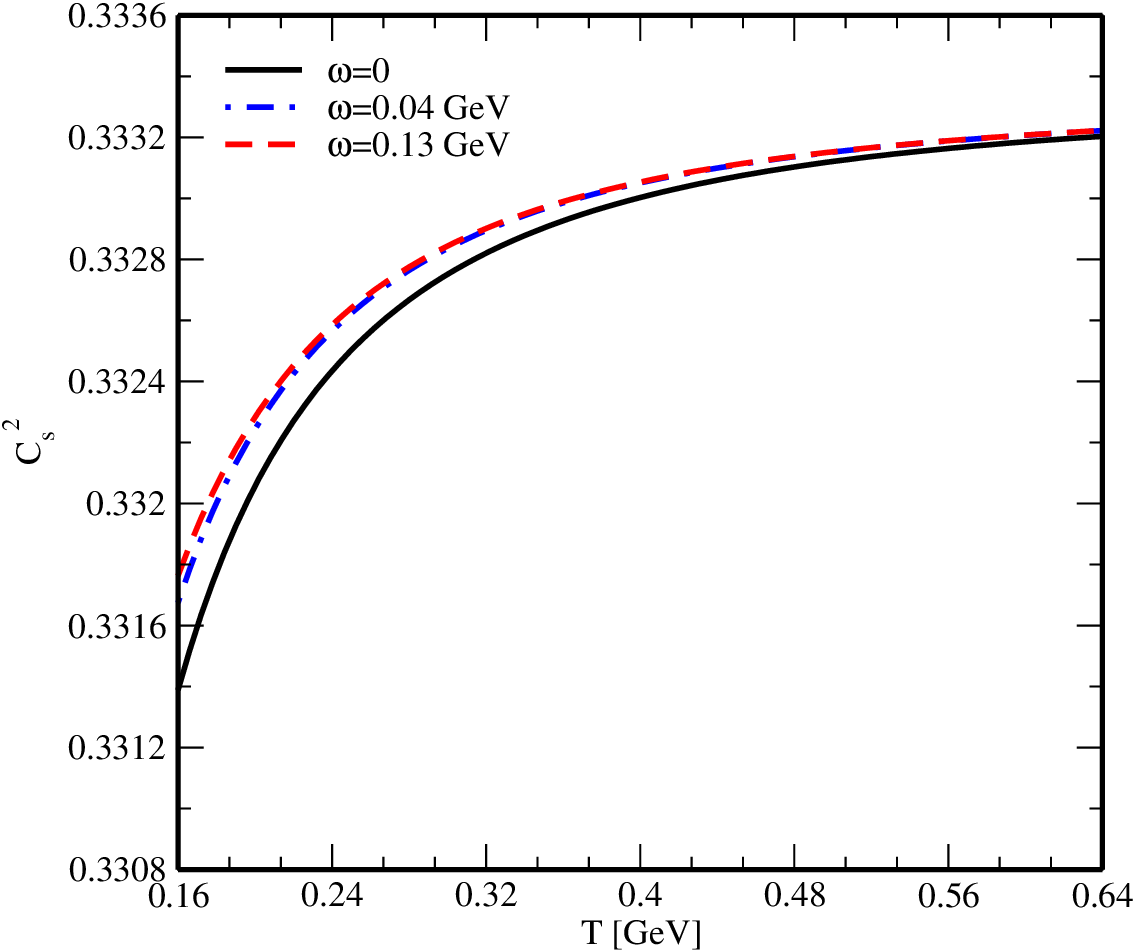}&
\hspace{0.73 cm}
\includegraphics[width=7.4cm]{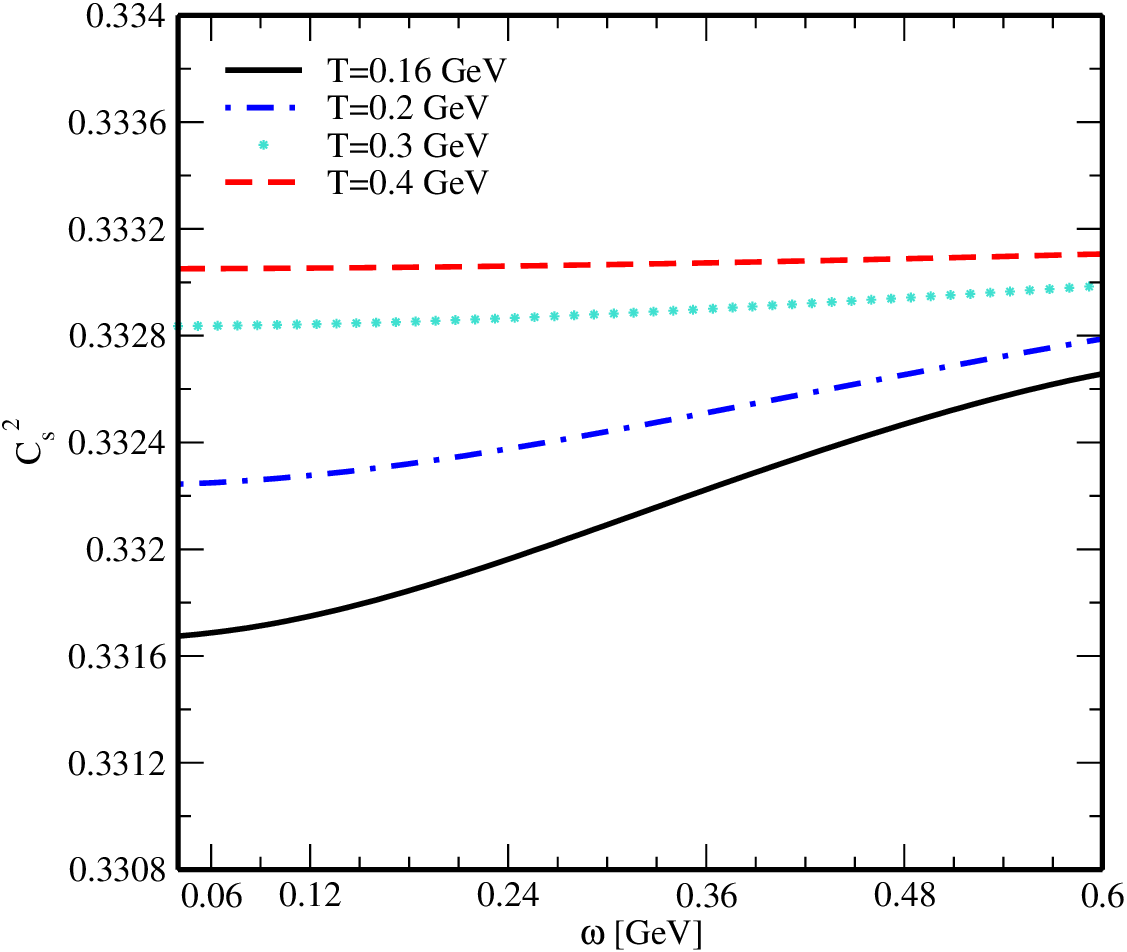} \\
\hspace{9.1mm}a & \hspace{18.86mm}b
\end{tabular}
\caption{Variation of the squared speed of sound: (a) as a function of temperature for different angular velocities and (b) as a function of angular velocity for different temperatures.}\label{Fig.7}
\end{center}
\end{figure}

The hydrodynamic expansion of matter is in general given by the conservation of the energy-momentum tensor, 
\begin{equation}\label{C.E.}
\partial_\mu T^{\mu \nu}=0
~.\end{equation}
In the absence of viscous forces, the energy-momentum tensor is written as
\begin{equation}\label{tmu}
T^{\mu\nu}= (\varepsilon+P)u^\mu u^\nu - g^{\mu \nu} P 
~.\end{equation}
The hydrodynamic evolution preserves boost invariance, which in ultrarelativistic heavy-ion collisions is expected to be realized near mid-rapidity.
Thus, the energy-momentum conservation equation \eqref{C.E.} gets simplified into 
\begin{equation}\label{E.M.}
\frac{d \varepsilon}{d \tau} = -\frac{\varepsilon+P}{\tau}
~.\end{equation}
The solution of the hydrodynamic equation of motion \eqref{E.M.} can be written as
\begin{eqnarray}
\label{eq0}
\varepsilon(\tau) \tau^{1+c_s^2}= \varepsilon(\tau_i)\tau_i^{1+c_s^2}={\rm constant}
~,\end{eqnarray}
where its dependence on the equation of state can be realized through the speed of sound ($c_s$). In the above equation, $\tau_i$ denotes the initial proper time and $\varepsilon (\tau_i)$ represents the initial energy density. By considering the energy density $\varepsilon \sim T^4$ and assuming the ideal relativistic gas limit for the squared speed of sound, $c_s^2=\frac{1}{3}$ in the above equation, one gets the cooling law as
\begin{eqnarray}
\label{eq1}
T^3 \tau =T_i^3 \tau_i
~.\end{eqnarray}
Here, $T_i$ is the initial temperature. In the presence of dissipative forces, the 
energy-momentum tensor translates into 
\begin{equation}
T^{\mu\nu}= (\varepsilon+P)u^\mu u^\nu - g^{\mu \nu} P +\Pi^{\mu\nu}
\label{tmun1}
,\end{equation}
where $\Pi^{\mu \nu}$ is the viscous tensor in a first-order theory, with the following form, 
\be
\Pi^{\mu\nu}=\eta \langle \nabla^\mu u^\nu \rangle+\zeta\Delta^{\mu\nu}\nabla\cdot u
~.\ee
Here, $\nabla^\mu=(g^{\mu \nu} - u^\mu u^\nu) \partial_\nu$. In a first-order theory, $T^{\mu\nu}$ can 
contain terms up to first order in the gradients of fluid degrees of freedom. Bulk viscosity has not 
been incorporated into our analysis of hydrodynamic expansion because it is not developed in the early stages of expansion and its value is negligibly small for massless flavors \cite{Weinberg:BOOK'1972,Muronga:PRL88'2002}. For a boost-invariant expansion in first-order dissipative hydrodynamics, the equation of motion is given \cite{Israel:1979wp,Baier:2006um} by
\be
\frac{d \varepsilon} {d \tau}+\frac{\varepsilon+P}{\tau}=\frac{4\eta}{3\tau^2}
~.\label{eqbj2}
\ee
The solution of the above equation can be decomposed into the nonviscous and the viscous components as
\begin{eqnarray}
\label{eqs1}
\nonumber\varepsilon(\tau) \tau^{1+c_s^2}+\left(\frac{4a_f T_{i}^3 \tau_{i}}{3}\right)\left(\frac{\eta}{s}\right)\frac{\tau^{1+c_s^2}}{{\tilde{\tau}}^2} &=& \varepsilon(\tau_i)\tau_i^{1+c_s^2}
+\left(\frac{4a_f T_{i}^3 \tau_{i}}{3}\right)
\left(\frac{\eta}{s}\right)\frac{\tau_i^{1+c_s^2}}{ {\tilde{\tau_i}}^2} \\ &=& {\rm constant}
~,\end{eqnarray}
where $a_f=(16+21N_f/2)\pi^2/90$, ${\tilde{\tau}}^2=(1-c_s^2)\tau^2$ and ${\tilde{\tau}}_i^2=(1-c_s^2)\tau_i^2$. The first terms on both sides of eq. \eqref{eqs1} resemble the ideal case, 
while the second terms express the first order viscous corrections. In the first-order viscous 
hydrodynamics, the evolution of the temperature with the proper time ($\tau$) is expressed \cite{Hosoya:NPB250'1985,Kouno:PRD41'1990} as
\begin{equation}
T(\tau)=T_i\left(\frac{\tau_i}{\tau}\right)^{1/3}\left[1+\frac{2\eta}{3\tau_i{T_i}s}\left\lbrace1-\left(\frac{\tau_i}{\tau}\right)^{2/3}\right\rbrace\right]
,\end{equation}
where the first term corresponds to the ideal hydrodynamic case and the second term is the first order viscous correction, involving the ratio $\eta/s$. It would be interesting to see how the solutions \eqref{eq0} and \eqref{eqs1} are influenced by the rotation of the strongly interacting matter. In eq. \eqref{eqs1}, the squared speed of sound and the ratio $\eta/s$ encode the influence of rotation on the evolution of the energy density. Thus, the effect of a rotating QGP medium on the expansion dynamics can be inferred by the thermodynamic equation of state and the viscosity. Figure \ref{Fig.7} 
shows the squared speed of sound ($c_s^2$) as a function of temperature for different angular velocities and as a function of angular velocity for different temperatures. The results show that $c_s^2$ increases with angular velocity and the impact of rotation on the equation of state is more pronounced in the low temperature regime, whereas at higher temperatures, the effect of rotation becomes minimal. 

\begin{figure}[]
\begin{center}
\includegraphics[width=7.4cm]{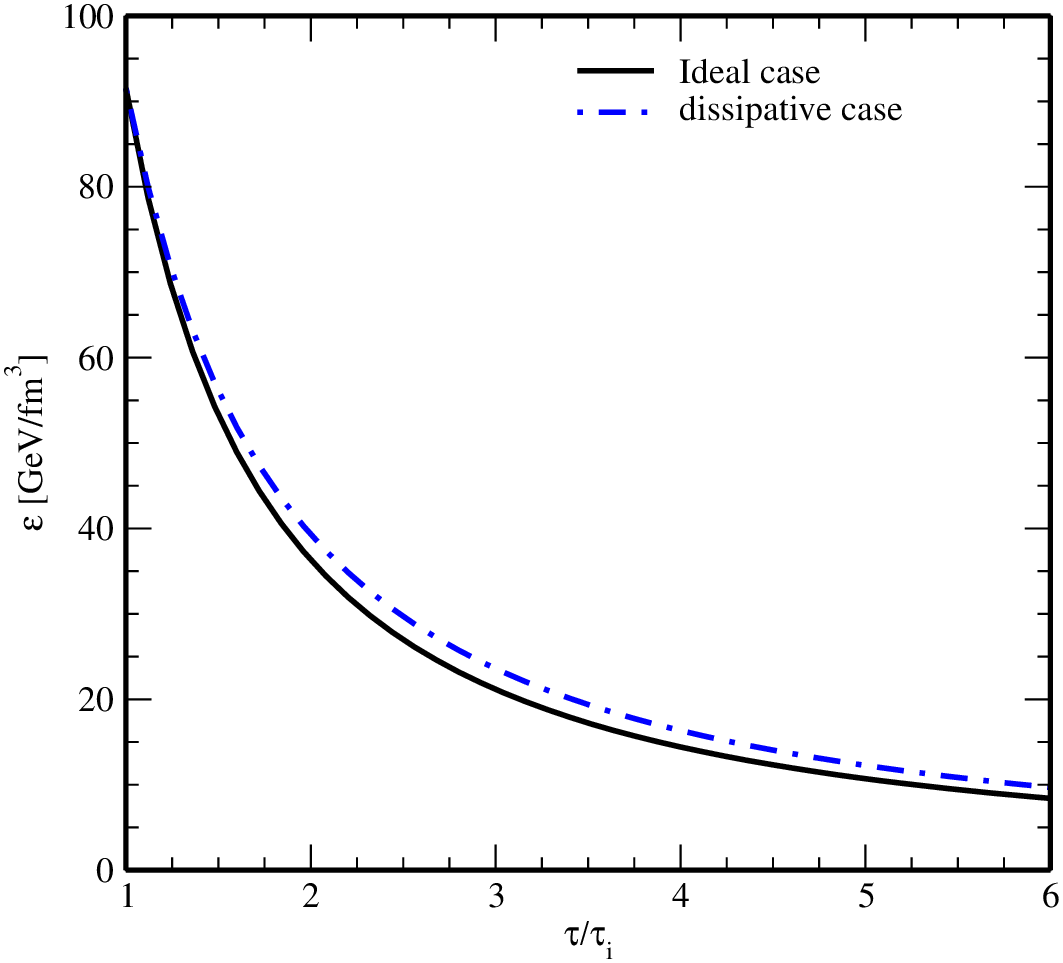}
\caption{Evolution of the energy density without rotation in ideal and dissipative cases.}\label{Fig.8}
\end{center}
\end{figure}

\begin{figure}[]
\begin{center}
\begin{tabular}{c c}
\includegraphics[width=7.4cm]{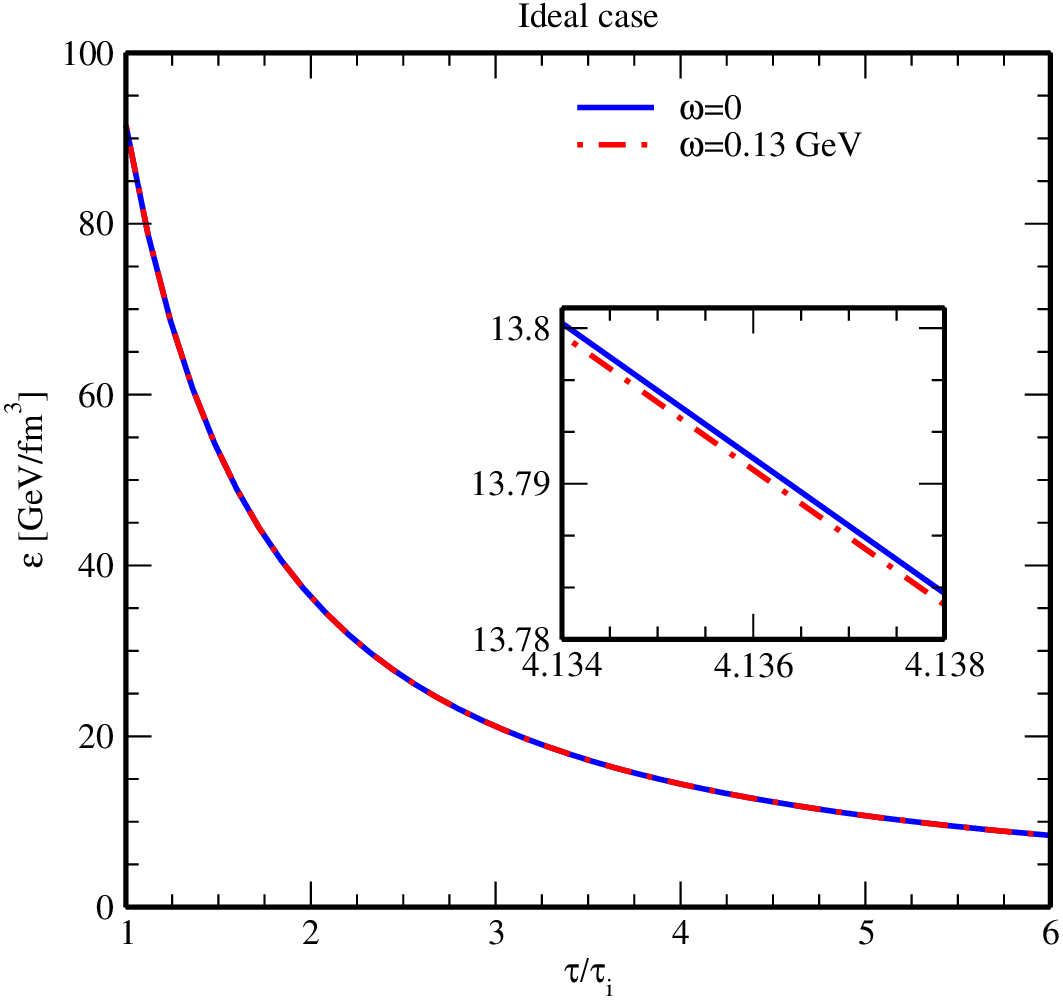}&
\hspace{0.73 cm}
\includegraphics[width=7.4cm]{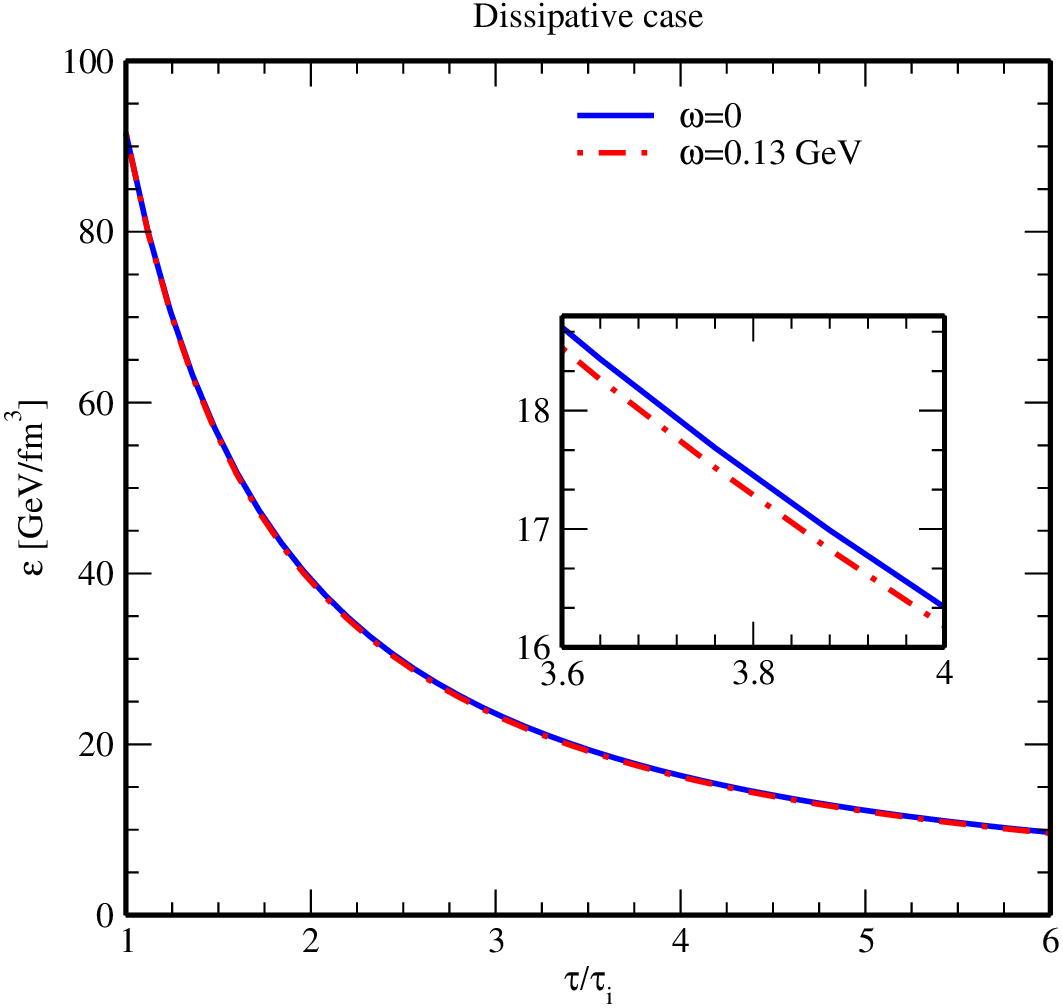} \\
\hspace{8mm}a & \hspace{17.1mm}b
\end{tabular}
\caption{Evolution of the energy density: (a) in the ideal case and (b) in the dissipative case for different angular velocities.}\label{Fig.9}
\end{center}
\end{figure}

\begin{figure}[]
\begin{center}
\begin{tabular}{c c}
\includegraphics[width=7.4cm]{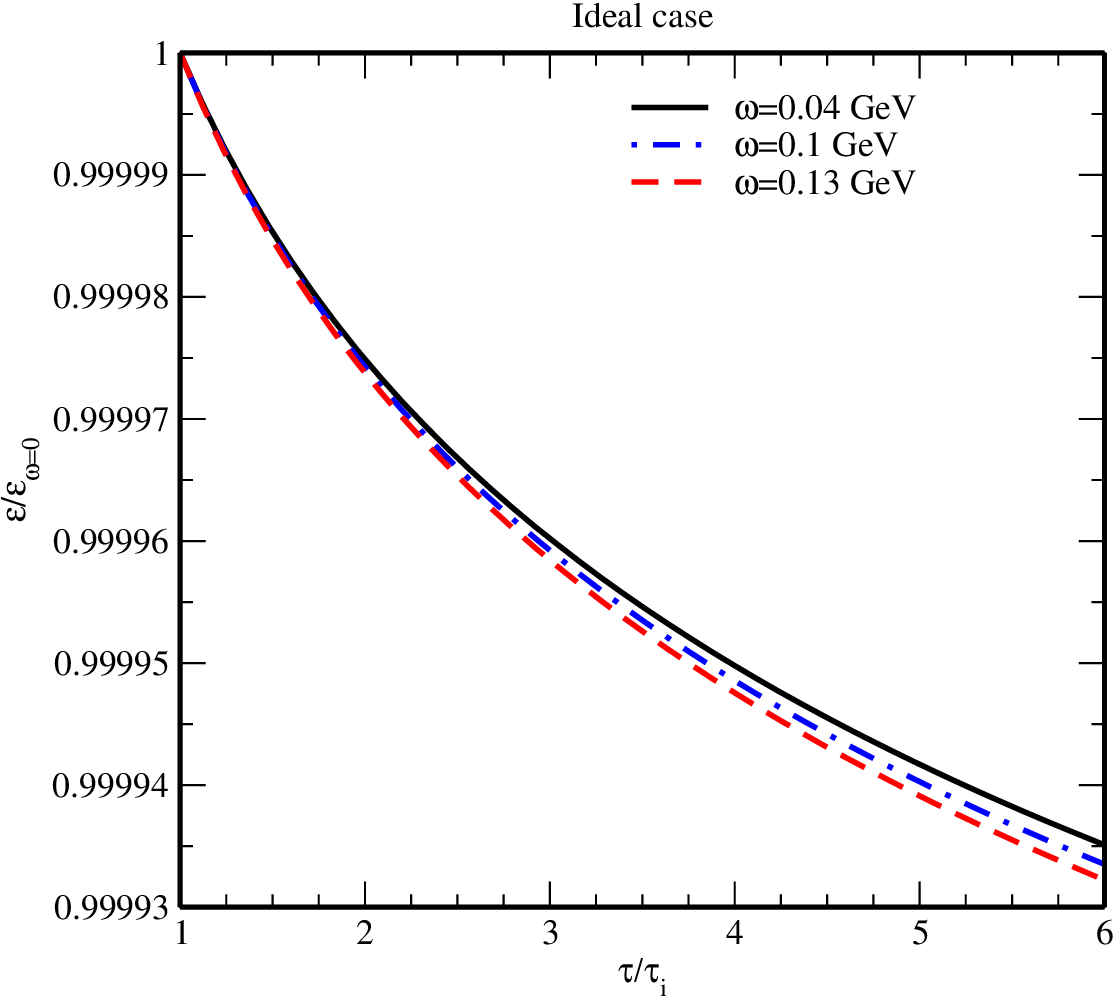}&
\hspace{0.73 cm}
\includegraphics[width=7.4cm]{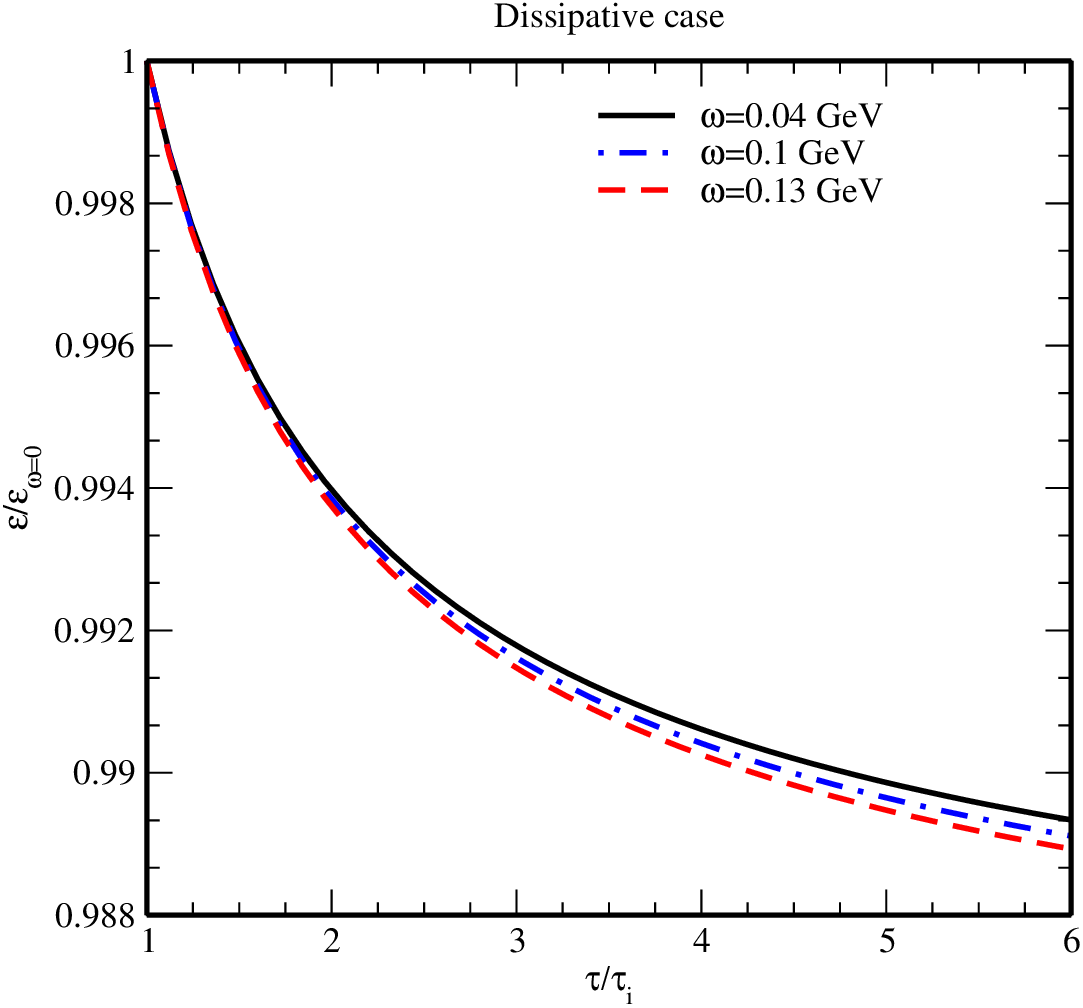} \\
\hspace{8mm}a & \hspace{17.1mm}b
\end{tabular}
\caption{Evolution of the energy density, scaled by its value in the nonrotating medium, for different angular velocities (a) in the ideal case and (b) in the dissipative case.}\label{Fig.10}
\end{center}
\end{figure}

In the study of the expansion dynamics of matter, we set $\tau_i=0.25$ fm/c and $T_i=0.46$ GeV for the centrality $(20 - 30)$\% at a center-of-mass energy of $\sqrt{s_{NN}}=200$ GeV, based on RHIC Au$+$Au 
data \cite{Kisiel:PRC79'2009}. The initial energy density, $\varepsilon_i=91.5176$ GeV/${\rm{fm}}^3$ is calculated from the equation of state. To understand how rotation affects the hydrodynamic expansion, we have explored the evolution of the energy density as a function of the proper time, both in the absence and in the presence of angular velocity, for both ideal and dissipative fluids (figures \ref{Fig.8}, \ref{Fig.9} 
and \ref{Fig.10}). It is found that the energy density decays more slowly in the dissipative fluid compared to the ideal fluid (figure \ref{Fig.8}). However, within ideal hydrodynamics, energy density decays faster for rotating matter compared to the stationary case (figure \ref{Fig.9}a). Similarly, faster decay of energy density for rotating matter compared to the stationary case is observed within dissipative hydrodynamics (figure \ref{Fig.9}b). This behavior at finite rotation is consistent with our earlier observation on the equation of state (figure \ref{Fig.7}), where the speed of sound gets enhanced by the introduction of angular velocity. Figure \ref{Fig.10} shows that, as the angular velocity increases, the energy density deviates further from its nonrotating counterpart in the ideal case (figure \ref{Fig.10}a) as well as in the dissipative 
case (figure \ref{Fig.10}b). Thus, the medium cools off faster as the rotation increases. In the ideal case, rotation increases the expansion rate and redistributes internal energy into flow, producing faster decay of energy density in the rotating fluid than in the stationary fluid. In the dissipative case, dissipative stresses couple strongly to the velocity gradients of a rotating fluid. The large velocity gradients and expansion rates in the rotating system generate more entropy density and enhance the decay rate of energy density compared to the stationary system. The faster decay of energy density in rotating matter can be further understood from the fact that the rotation redistributes energy into rotational motion and 
modifies the thermodynamic distribution functions. These effects increase the effective expansion rate and contribute in decaying the energy density more rapidly in a rotating system than in a stationary system. 

\section{Conclusions}
In this work, we studied how the rotation of the QGP medium affects its viscous properties and Bjorken expansion. The shear ($\eta$) and bulk ($\zeta$) viscosities were calculated 
using the novel relaxation time approximation within the kinetic theory approach, incorporating finite angular velocity. Particle interactions were taken into account through their quasiparticle masses. We also explored observable effects associated with the aforesaid viscosities, such as the flow characteristic, the specific shear viscosity and the specific bulk viscosity. Our results indicate that the rotation enhances both $\eta$ and $\zeta$, confirming that the rotation of the medium enhances momentum transfer and local pressure fluctuations. Additionally, the viscous behavior of the QGP medium gets suppressed at finite angular velocity due to the increase in the Reynolds number, confirming that the flow remains laminar even for a rotating medium. Furthermore, since the ratio $\eta/s$ approaches the conjectured lower bound at finite angular velocity, the rotating medium exhibits nearly perfect fluid behavior. The behavior of the ratio $\zeta/s$ suggests that, rotation drives the medium away from conformal symmetry. Finally, our findings also show that, a rotating medium evolves faster than a stationary one. 

\section{Acknowledgments}
One of the authors (S. R.) acknowledges financial support from the ANID Fondecyt Postdoctoral Grant 3240349 
for this work. N. N. acknowledges support from ANID (Chile) FONDECYT Iniciaci\'on Grant No. 11230879.

\end{document}